\documentclass{llncs}
\usepackage{amssymb}
\usepackage{comment}
\title{Cliquewidth and Knowledge Compilation}
\author{Igor Razgon \thanks{I would like to thank Fedor Fomin for his help in shaping
of my understanding of the structural graph parameters.}\inst{1} \and Justyna Petke \inst{2}}
\institute{
Department of Computer Science and Information Systems, \\ 
Birkbeck, University of London \email{igor@dcs.bbk.ac.uk} \and
Department of Computer Science, \\ University College London
\email{J.Petke@cs.ucl.ac.uk} 
} 
\begin{document}
\maketitle
\begin{abstract}
In this paper we study the role of cliquewidth in succinct representation of Boolean
functions. Our main statement is the following: Let $Z$ be a Boolean circuit having
cliquewidth $k$. Then there is another circuit $Z^*$ computing the same function as $Z$
having treewidth at most $18k+2$ and which has at most $4|Z|$ gates where $|Z|$ is the number of gates of $Z$. 
In this sense, cliquewidth is not more `powerful' than
treewidth for the purpose of representation of Boolean functions. We believe this is quite 
a surprising fact because it contrasts the situation with graphs where an upper bound on the 
treewidth implies an upper bound on the cliquewidth but not vice versa.

We demonstrate the usefulness of the new theorem for knowledge compilation.
In particular, we show that a circuit $Z$ of cliquewidth $k$ can be compiled into a Decomposable
Negation Normal Form ({\sc dnnf}) of size $O(9^{18k}k^2|Z|)$ and the same runtime. To the best of our knowledge, 
this is the first result on efficient knowledge compilation parameterized by cliquewidth of a 
Boolean circuit. 
\end{abstract}
\section{Introduction}
Cliquewidth is a graph parameter, probably best known for its role in the design of fixed-parameter algorithms
for graph-theoretic problems \cite{CoMaRo}. In this context the most interesting property of cliquewidth is that it
is `stronger' than treewidth in the following sense: if all graphs in some (infinite) class have treewidth bounded by some
constant $c$, then the cliquewidth of the graphs of this class is also bounded by a constant $O(2^c)$. However, the
opposite is not true. Consider, for example, the class of all complete graphs. The treewidth of this class is unbounded
while the cliquewidth of any complete graph is $2$.

In this paper we essentially show that, roughly speaking, cliquewidth of a Boolean function is not
a stronger parameter than its treewidth. In particular, given a Boolean circuit $Z$, we define its cliquewidth
as the cliquewidth of the DAG of this circuit and the treewidth as the treewidth of the undirected graph 
underlying this DAG. The main theorem of this paper states that for any circuit $Z$ of cliquewidth $k$ there is
another circuit $Z^*$ computing the same function whose treewidth is at most $18k+2$ and the number of gates is at most
$4$ times the number of gates of $Z$. Moreover, if $Z$ is accompanied with the respective clique decomposition then
such a circuit $Z^*$ (and the tree decomposition of width $18k+2$) can be obtained in time $O(k^2n)$.
The definition of circuit treewidth is taken from \cite{OBDDTWJha} and the definition of circuit cliquewidth naturally
follows from the treewidth definition. In fact, the relationship between circuit treewidth and cliquewidth is 
put in \cite{OBDDTWJha} as an open question.

We demonstrate that the main theorem is useful for knowledge compilation, that is, compact representation of Boolean
functions that allows to efficiently answer certain queries regarding the considered function.
In particular, we show 
that any circuit $Z$ of cliquewidth $k$ can be compiled into decomposable negation normal form
({\sc dnnf}) \cite{DarwicheJACM} of size $O(9^{9k}k^2|Z|)$ (where $|Z|$ is the number of gates)
by an algorithm taking the same runtime. To the best of our knowledge, this is the first result 
on space-efficient knowledge compilation parameterized by cliquewidth. We believe this
result is interesting because the parameterization by
cliqewidth, compared to treewidth, allows to capture a wider class of inputs including those circuits
whose underlying graphs are dense. 

This bound is obtained as an immediate corollary of the main theorem
and the $O(9^{t}t^2|Z|)$ bound on the {\sc dnnf} size for the given circuit $Z$, where $t$ is the treewidth
of $Z$. The intermediate step for the latter result is an $O(3^p(|C|+n))$ bound of the {\sc dnnf} 
size of the given {\sc cnf} where $C$ and $n$ are, respectively the number of clauses and variables
of this {\sc cnf} and $p$ is the treewidth of its \emph{incidence} graph. All these $3$ bounds significantly 
extend the currently existing bound  $O(2^rn)$ of \cite{DarwicheJACM} where $r$ is the treewidth of 
the \emph{primal} graph of the given {\sc cnf}. For example, if the given {\sc cnf} has large clauses (and hence
a large treewidth of the primal graph) then the $O(2^rn)$ bound becomes practically
infeasible while the $O(3^p(C+n))$ bound may be still feasible provided a small treewidth of the incidence
graph and a number of clauses polynomially dependent on $n$.  
\section{Related Work}
The algorithmic power of cliquewidth stems from the meta-theorem of \cite{CoMaRo}
stating that any problem definable in Monadic Second Order Logic (MSO$_1$) can be solved in linear time for a class
of graphs of fixed cliquewidth $k$. The cliquewidth of the given graph is NP-hard to compute \cite{CWNP}
and it is not known to be FPT. On the other hand, cliquewidth is FPT approximable by an FPT computable
parameter called \emph{rankwidth} \cite{CWDApprox,RWDCompute}. As said above, there are classes of graphs with unrestricted treewidth
and bounded cliquewidth. However, it has been shown in \cite{CWDNoBicliques} that the only reason for treewidth to be much larger
than cliquewidth is the presence of a large complete bipartite graph (biclique) in the considered graph. In fact, we prove the main theorem of this 
paper by applying a transformation that eliminates all bicliques from the DAG of the given circuit. 


{\sc dnnf}s have been introduced as a knowledge compilation formalism in \cite{DarwicheJACM}, where it has been shown
that any {\sc cnf} on $n$ variables of treewidth $t$ of the primary graph can be compiled into a {\sc dnnf} of size $O(2^ttn)$
with the same runtime. A detailed analysis of special cases of {\sc dnnf} has been provided in \cite{DerMar}. In particular,
it has been shown that Free Binary Decision Diagrams ({\sc fbdd}) and hence Ordered Binary Decision Diagrams ({\sc obdd}) can be seen as special cases of {\sc dnnf}. 
In fact, there is a separation between {\sc dnnf} and {\sc fbdd} \cite{SepDNNFFBDD}. This additional expression power of {\sc dnnf} has 
its disadvantages: a number of queries that can be answered 
in polynomial time (polytime) for {\sc fbdd} and {\sc obdd} are NP-complete for {\sc dnnf} \cite{DerMar}. This trade-off led to investigation of subclasses
of {\sc dnnf} that, on one hand, retain the succinctness of {\sc dnnf} for {\sc cnf}s of small treewidth and, on the other hand, have an increased
set of queries that can be answered in polytime. Probably the most notable result obtained in this direction are
Sentential Decision Diagrams ({\sc sdd}) \cite{SDD} that, on one hand, can answer in polytime the equivalence query
(possibility to answer this query in polytime for {\sc obdd}s is probably the main reason why this formalism is very
popular in the area of verification) and, on the other hand, retains the same upper bound dependence on treewidth
as {\sc dnnf}. 

In fact the size of {\sc obdd} can also be efficiently parameterized by the treewidth of the initial representation of the considered function. 
Indeed, there is an {\sc obdd} of size $O(n2^p)$ where $p$ is the pathwidth of the primal graph of the given
{\sc cnf} and of size $(n^{O(t)})$ where $t$ is the treewidth of the graph, see e.g. \cite{VardiTWD}.
It is shown in \cite{OBDDTWJha} that similar pattern retains if we consider the pathwidth and treewidth of a circuit but in the former
case $p$ is replaced by an exponential function of $p$ and in the latter case, $t$ is replaced by a double exponential function
of $t$. 
   
\section{Preliminaries}
A \emph{labeled} graph $G=(V,E,{\bf S})$, in addition to the usual set $V(G)$ of vertices and a set $E(G)$ of edges, contains
a component ${\bf S}(G)$, which is a partition of $V(G)$. Each element of the partition class is called a \emph{label}.
A \emph{simplified clique decomposition} ({\sc scd}) is a pair $(T,{\bf G})$ where $T$ is a rooted tree and ${\bf G}$ is a family of labeled graphs.
Each node $t$ of $T$ is associated with a graph $G(t)$, which is defined as follows.
If $t$ is a leaf node, then $G(t)=(\{v\},\emptyset,\{\{v\}\})$. Assume that $t$ has two children $t_1$ and $t_2$ and let $G_1=G(t_1)$
and $G_2=G(t_2)$.
Then $V(G_1) \cap V(G_2)=\emptyset$ and $G(t)=(V(G_1) \cup V(G_2), E(G_1) \cup E(G_2), {\bf S}(G_1) \cup {\bf S}(G_2))$.
Finally, assume that $t$ has only one child $t_1$ and let $G_1=G(t_1)$. Graph $G(t)$ can be obtained from $G_1$ by one
of the following three operations:
\begin{itemize}
\item {\bf Adding a new vertex}. There is $v \notin V(G_1)$ such that $G(t)=(V(G_1) \cup \{v\},E(G_1),{\bf S}(G_1) \cup \{\{v\}\})$.
\item {\bf Union of labels}. There are $S_1,S_2 \in {\bf S}(G_1)$ such that $G(t)=(V(G_1),E(G_1),$ $({\bf S}(G_1) \setminus \{S_1,S_2\}) \cup \{S\})$.
We say that $S_1$ and $S_2$ are \emph{children} of $S$.
\item {\bf New adjacency}. There are $S_1,S_2 \in {\bf S}(G_1)$ such that $G(t)=(V(G_1),E(G_1) \cup \{\{u,v\}|u \in S_1,v \in S_2\},{\bf S}(G_1))$.
We say that $S_1$ and $S_2$ are \emph{adjacent}. 
\end{itemize}

The width of a node $t$ of $T$ is $|{\bf S}(G_t)|$. The width of $(T,{\bf G})$ is the largest width of a node $t$ of $T$.
Let $r$ be the root of $T$. Then we say that $(T,{\bf G})$ is an {\sc scd} of $G(r)$ and of $(V(G(r)),E(G(r)))$
(the unlabeled version of $G(r)$. The \emph{simplified cliquewidth} ({\sc scw}) of a graph $G$ is the smallest width among all 
{\sc scd}s of $G$. The definition of {\sc scd} is closely related to the standard notion of clique decomposition. 
In fact {\sc scw} of a graph $G$ is at most twice larger than the cliquewidth of $G$. The details of comparison are postponed
to the appendix.

Clique decomposition and {\sc scd} are easily extended to the directed case.
In fact the notion of cliquewidth has been initially proposed for the directed case, as noted in  \cite{COMMA2010}. 
The only change is that the new adjacency operation adds to $G(t)$ all possible directed arcs from label $S_1$ to
label $S_2$ instead of undirected edges. In this case we say that there is an arc from $S_1$ to $S_2$.

We denote $\bigcup_{t \in V(T)}{\bf S}(G(t))$ by ${\bf S}={\bf S}(T,{\bf G})$ and call it \emph{the set of labels} of $(T,{\bf G})$.



A tree decomposition of a graph $G$ is a pair $(T,{\bf B})$ where $T$ is a tree and the elements of ${\bf B}$ are subsets of vertices
called \emph{bags}. There is a mapping between the nodes of $T$ and elements of ${\bf B}$. 
Let us say a vertex $v$ of $G$ is \emph{contained} in a node $t$ of $T$ if $v$ belongs to the bag $B(t)$ of $t$.
Two properties of a tree decomposition are \emph{connectedness} (all the nodes containing the given vertex
$v$ form a subtree of $T$), \emph{adjacency} (each edge $\{u,v\}$ is a subset of some bag), and \emph{union} (the union
of all bags is  $V(G)$. 
In this paper we consider the treewidth of a directed graph as the treewidth of the underlying undirected graph.

Boolean circuits considered in this paper are over the basis $\{\vee,\wedge,\neg\}$.
In such a circuit there are input gates (having only output wires) corresponding to variables and constants
$true$ and $false$. The output of each gate of a circuit $Z$ computes a function on the set of input variables.
We denote by $functions(Z)$ the set of all functions computed by the gates of $Z$.
The number of gates of $Z$ is denoted by $|Z|$.

A clique or tree decomposition of a circuit $Z$ is the respective decomposition of the 
DAG of $Z$. In our discussion, we often associate the vertices of the DAG with the respective gates.
\emph{De Morgan circuits} are a subclass of circuits where the inputs of all the {\sc not} gates are variables
(i.e. the outputs of {\sc not} gates serve as negative literals). For a gate $g$ of $Z$, denote by $Var(g)$
the set of variables having a path to $g$ in the DAG of $Z$. A circuit $Z$ has the \emph{decomposability} property
if for any two in-neighbors $g_1$ and $g_2$ of an {\sc and} gate $g$, $Var(g_1) \cap Var(g_2)=\emptyset$.
{\sc dnnf} is a decomposable De Morgan circuit. 
When we consider a general circuit $Z$, we assume that it does not
have constant input gates, since these gates can be propagated by removal of some gates of $Z$, which in turn
does not increase the cliquewidth nor the treewidth of the circuit. However, for convenience of reasoning, 
we may use constant input gates when we describe construction of a {\sc dnnf}. If the given circuit $Z$ is a {\sc cnf}
then its variables-clauses relation can be represented by the \emph{incidence graph}, 
a bipartite graph with parts corresponding to variables and clauses and a variable-clause
edge representing occurrence of a variable in a clause. 

\section{From small cliquewidth to small treewidth}
The central result of this section is the following theorem:

\begin{theorem} \label{finalwidth}
Let $F$ be a circuit of cliquewidth $k$ over $n$ variables
Then there is a circuit $F^*$ of treewidth at most $18k+2$ and $|F^*| \leq 4|F|$
such that $functions(F) \subseteq functions (F^*)$. Moreover, 
given $F$ and a clique decomposition of $F$ of width $k$ there is
an $O(k^2n)$ algorithm constructing $F^*$ and a tree decomposition of
$F^*$ of width at most $18k+2$ having at most $2|F|$ bags.
\end{theorem}

The rest of this section is the proof of Theorem \ref{finalwidth}.
The main idea of the proof is to replace `parts' of the given circuit forming large
bicliques by circuits computing equivalent functions where such bicliques do not occur.
As an example consider a {\sc cnf} of $3$ clauses $C_1=(a_1 \vee a_2 \vee a_3 \vee b_1)$,
$C_2=(a_2 \vee a_2 \vee a_3 \vee b_2)$ and $C_3=(a_1 \vee a_2 \vee a_3 \vee b_3)$. 
The circuit of this graph contains a biclique of order $3$ created by $C_1,C_2,C_3$
on one side and $a_1,a_2,a_3$ on the other one. This biclique can be eliminated 
by the introduction of additional {\sc or} gate $C_4$ having input $a_1,a_2,a_3$ and output $C_4$ 
so that the clauses $C_1,C_2,C_3$ are transformed into $(b_1 \vee c_4),(b_2 \vee c_4),(b_3 \vee c_4)$, 
respectively. It is not hard to see that the new circuit computes the same function as the original 
one. This is the main idea behind the construction of circuit $F^*$. The formal description
of the construction is given below.

For the purpose of construction of $F^*$
we consider a type respecting {\sc scd} $(T,{\bf G})$ of $F$ where each non-singleton label is one of the following:
\begin{itemize}
\item A \emph{unary} label containing input gates and negation gates.
\item An \emph{{\sc and}} label containing {\sc and} gates.
\item An \emph{{\sc or}} label containing {\sc or} gates.
\end{itemize}
 
The following lemma essentially follows from splitting each label of the given clique decomposition into
three type respecting labels. 

\begin{lemma} \label{typerespect}
Let $k$ be the cliquewidth of $F$ and let $k^*$ be the smallest width of an {\sc scd} of $F$
that respects types. Then $k^* \leq 6k$.
\end{lemma}

{\bf Proof.}
Let $(T^*, {\bf G^*})$ be an {\sc scd} of $Z$ having width at most $2k$ (existing since
the cliquewidth is $k$).
In each graph $G' \in {\bf G^*}$ split each label into at most $3$ subsets so that
each subset contains one type of the gates as specified above. Clearly, the resulting
number of labels is at most $3$ times larger than the original one. The resulting
structure is not necessarily an {\sc scd}. In particular, there may
be situation when the graph associated with a node is the same as the graph associated 
with the parent node because the union operation in the parent has been reversed by
the splitting. Also, the new adjacency operation may become applied between more than
one pair of labels. However, a legal {\sc scd} is easy to recover:
the 'redundant' parent nodes can be removed (since they are unary this will no
cause problems with the structure of the binary tree) and each node with
a multiple adjacency operation can be replaced by a sequence of nodes applying these
operations one by one. $\blacksquare$

Given a type respecting {\sc scd} $(T,{\bf G})$, let us construct the circuit $F^*$. 
In the first stage, we associate each label $S \in {\bf S}$ with a set of gates as follows:
\begin{itemize}
\item If $S$ is non-singleton then it is associated with an {\sc and} gate denoted by $oand(S)$
and an {\sc or} gate denoted by $oor(S)$.
\item If $S$ is non-singleton and does not contain input gates then it is associated with an additional
gate called $in(S)$ whose type is determined as follows: If $S$ is an {\sc and} or {\sc or} label
then $in(S)$ is  an {\sc and} or {\sc or} gate, respectively. If $S$ is a unary label then
$in(S)$ is a circuit (perceived as a single atomic gate) consisting of two {\sc not} gates, the output of
one of them is the input of the other. So, the input of the former and the output of the latter are,
respectively, the input and output of $in(S)$.
\item Each singleton label $\{g\}$ is associated with the gate $g$
of $F$. We call the gates associated with singleton labels \emph{original gates} because
they are the gates of $F^*$ appearing in $F$. For the sake of uniformity,
for each original gate $g$ associated with label $S$, we put $g=oand(S)=oor(S)=in(S)$.
\end{itemize}

The wires of $F^*$ are described below. When we say that
there is a wire from gate $g_1$ to gate $g_2$, we mean that the wire is
\emph{from the output} of $g_1$ \emph{to the input} of $g_2$.
\begin{itemize}
\item {\bf Child-parent wires.}
Let $S_1$ and $S_2$ be labels of $(T,{\bf G})$ such that
$S_1$ is a child of $S_2$. Then there is a wire from $oand(S_1)$ to $oand(S_2)$ and a wire from $oor(S_1)$ to $oor(S_2)$.
\item {\bf Parent-child wires.}
Let $S_1$ and $S_2$ be as above and assume that $S_2$ does not contain input gates.
Then there is a wire from $in(S_2)$ to $in(S_1)$. That is, the direction of child-parent wires
is opposite to the direction of parent-child wires. \footnote{We would like to thank the anonymous
referee, for helping us to identify a typo in this definition that occurred in the first version of
the manuscript.}
\item {\bf Adjacency wires.} Assume that in $(T,{\bf G})$ there is an arc from $S_1$ to $S_2$ (established by the new adjacency node). Then the following cases apply:
\begin{itemize}
\item If $S_2$ is an {\sc and} label then put a wire from $oand(S_1)$ to $in(S_2)$.
\item If $S_2$ is an {\sc or} label then put a wire from $oor(S_1)$ to $in(S_2)$.
\item If $S_2$ is a unary label consisting of negation gates only then put a wire from
an arbitrary one of $oand(S_1)$ or $oor(S_1)$ to $in(S_2)$.
\end{itemize}
\end{itemize}
 
Finally, we remove $in(S)$ gates that have no inputs.
This removal may be iterative as removal of one gate may leave without input another one.


It is not hard to see by construction that $F$ and $F^*$ have the same input gates.
This gives us possibility to state the following theorem with proof in Section \ref{samef}.

\begin{theorem} \label{samefunctions}
$F^*$ is a well formed circuit.
The output of each original gate $g$ of $F^*$  computes exactly the same function (in terms of input gates)
 as in $F$. 
\end{theorem}

In Section \ref{twidth}, we prove that the treewidth of $F^*$ is not much larger than the width of
$(T,{\bf G})$.

\begin{theorem} \label{widths}
There is a tree decomposition of $F^*$ with at most $2|F|$ bags having width at most $3k+2$, where $k$ is the width of $(T,{\bf G})$. 
\end{theorem}

Now we are ready to prove Theorem \ref{finalwidth}.

{\bf Proof of Theorem \ref{finalwidth}} 
Due to Theorem \ref{samefunctions}, $functions(F) \subseteq functions(F^*)$. If we take $(T,{\bf G})$ to be of the smallest possible
type respecting width then the treewidth of $F^*$ is at most $18k+2$ by combination of Theorem \ref{widths} and Lemma \ref{typerespect}.

To compute the number of gates of $F^*$, let $n$ be the number of gates of $F$, which is also the number of
singleton labels of $(T,{\bf G})$. Since each non-singleton label has two children (i.e. in the respective
tree of labels each non-leaf node is binary), the number of non-singleton labels is at most $n-1$.
By construction, $F^*$ has one gate per singleton label plus at most $3$ gates per non-singleton label,
which adds up to at most $4n$.

The technical details of the runtime derivation are postponed to the appendix.
$\blacksquare$

\subsection{Proof of Theorem \ref{samefunctions}} \label{samef}
We start with establishing simple combinatorial properties of $F^*$ (Lemmas \ref{childparent},\ref{andconnect},
\ref{notconnect},\ref{ubconnect}). 
A \emph{path} in a circuit is a sequence of gates so that the output of every gate (except the last one) is connected by a wire to the input of its successor. 
Let us call a path a \emph{connecting} path if it contains exactly one adjacency circuit.

\begin{lemma} \label{childparent}

\begin{itemize}
\item Any path $P$ of $F^*$ starting at an original gate and not containing adjacency wires contains
child-parent wires only.
\item Any path $P$ of $F^*$ ending at an original gate and not containing adjacency wires contains
parent-child wires only. 
\end{itemize}

\end{lemma}

{\bf Proof.}
The only possible wire to leave the original gate is a child-parent wire.
Any path starting from an original gate and containing child-parent wires only 
ends up in an $oand$ or $oor$ gate.  This means that the next wire (if not an adjacency one)
can be only another child-parent wire. Thus the correctness of the lemma for all
the paths of length $i$ implies its correctness for all such paths of length $i+1$,
confirming the first statement. 

For the second statement, we start from an original gate and go back \emph{against}
the direction of wires. The reasoning similar to the previous paragraph applies 
with the $in$ gates of non-singleton labels replacing the $oor$ and $oand$ ones.
$\blacksquare$  

\begin{lemma} \label{andconnect}
Let $g_1$ and $g_2$ be gates of $F$ such that $g_2$ is an {\sc and} or
an {\sc or} 
gate. Then there is a wire from $g_1$ to $g_2$ in $F$ if and only if
$F^*$ has a connecting path from $g_1$ to $g_2$ 
such that all the gates of this path except possibly $g_1$ are of the same type
as $g_2$.
\end{lemma}

{\bf Proof.}
We prove only the case where $g_2$ is an {\sc and} gate, the other case is symmetric. 
Let $P$ be a connecting path of $F^*$ from $g_1$ to $g_2$ of the specified kind. Let $g'_1$ and $g'_2$ be, respectively, the tail and the head gates of the adjacency wire. Then either 
$g_1=g'_1$ or the suffix of $P$ ending at $g'_1$ consists
of child-parent wires only according to Lemma \ref{childparent}.
It follows that $g'_1$ corresponds to a label containing $g_1$. Analogously, we conclude that either $g_2=g'_2$ or the suffix of $P$ starting at $g'_2$ contains only parent-child labels and hence the label corresponding to $g'_2$ contains $g_2$. Existence of the adjacency wire from the label of $g'_1$ to the label of $g'_2$ means that the {\sc scd} introduces all wires from the 
gates in the label of $g'_1$ to the gates in the label of $g'_2$. In particular, there is a wire from $g_1$ to $g_2$ in $F$. 

Conversely, assume that there is a wire from $g_1$ to $g_2$ in $F$. Then there are labels $S_1$ and $S_2$ containing 
$g_1$ and $g_2$, respectively, such that $(T,{\bf G})$ introduces an adjacency arc from $S_1$ to $S_2$. By construction of $F^*$ there is a gate $g'_1$ corresponding to $S_1$ and a gate $S'_2$ corresponding to $g_2$ such that $F^*$ has an adjacency wire from $g'_1$ to $g'_2$. Moreover, by the definition of a type respecting {\sc scd}, $S_2$ is
an {\sc and} label, hence $g'_2=in(S_2)$ is an {\sc and} gate. Furthermore, by construction of $F^*$
either $g_2=g'_2$ or there is a path from $g'_2$ to $g_2$ consisting of parent-child arcs only and {\sc and} gates only. Indeed, if $S_2$ is not a singleton then there is a wire from $in(S_2)$ to $in(S_3)$ containing $g_2$ since $S_3$ is the parent of $S_2$. Iterative application of this argument produces a path from $g'_2$ to $g_2$. Since $g_2$ is an {\sc and} gate, all gates in this path are 
{\sc and} gates by construction.  Thus the suffix exists. What about the prefix?
By construction, $g'_1=oand(S_1)$. Since $S_1$ contains $g_1$,
either $g'_1=g_1$ or there is a path from $g_1$ to $g'_1$ involving child-parent wires and {\sc and} gates only: just start at $g_1$ and go every time to the $oand$-gate of the parent
until $S_1$ has been reached. Thus we have established existence of the desired prefix.

It remains to be shown that the prefix and suffix do not intersect. However, this is impossible due to the disjointness of $S_1$ and $S_2$. 
$\blacksquare$


\begin{lemma} \label{notconnect}
Let $g_1$ and $g_2$ be the gates of $F$ such that $g_2$ is a {\sc not} gate. Then $F$ has a wire from $g_1$ to $g_2$ if and only if 
there is a connecting path $P$ in $F^*$ from $g_1$ to $g_2$ with the adjacency wire $(g'_1,g'_2)$ such that $g_1=g'_1$ and all the intermediate vertices in the suffix of $P$ 
starting from $g'_1$ are $in$-gates of unary labels containing negation gates only.
\end{lemma}

{\bf Proof.}
Let $P$ be a connecting path of $F^*$ of the specified form. Then either $g'_2=g_2$ or $g'_2$ corresponds to a label containing $g_2$. In both cases this means that $F$ has a wire from $g_1$ to $g_2$.

Conversely, assume that $F$ has a wire from $g_1$ to $g_2$.
Then there are labels $S_1$ and $S_2$ containing $g_1$ and $g_2$ such that $(T,{\bf G})$ sets an adjacency wire from $S_1$ to $S_2$. Observe that $S_1$ cannot contain more than one element because in this case $g_2$, a {\sc not} gate, will have two inputs. Furthermore, either $S_2$ contains $g_2$ only or $S_2$ is a unary label containing negation gates only
(because the input gates do not have input wires). In the latter case, the desired suffix from the head of the adjacency arc to $g_2$ follows by construction.  
$\blacksquare$

\begin{lemma} \label{ubconnect}
Any path of $F^*$ between two original gates that does not involve other original gates is a connecting path.
\end{lemma}

{\bf Proof.}
First of all, let us show that any path of $F^*$ between original gates involves at least one adjacency wire. 
Indeed, by Lemma \ref{childparent}, any path leaving an original gate and not having adjacency wires has only child-parent wires.
Such wires lead only to bigger and bigger labels and cannot end up with a singleton gate.
It follows that at least one adjacency wire is needed.

Let us show that additional adjacency wires cannot occur without original gates as intermediate vertices.
Indeed, the head of the first adjacency wire is an $in$ gate of some label $S$. Unless $S$ is a singleton,
the only wires leaving $in(S)$ are parent-child wires to the $in$ gates of the children of $S$. Applying this
argumentation iteratively, we observe that no other wires except parent-child wires are possible until
the path meets the $in$ gate of a singleton label. However, this is an original gate that cannot be 
an intermediate node in our path. It follows that any path between two original gates without other original 
cannot involve $2$ adjacency wires. Combining with the previous paragraph, it follows that any such path
involves exactly one adjacency wire, i.e. it is a connecting path.
$\blacksquare$ 

Using the lemmas above, it can be shown that any cycle in $F^*$ involves at least one original
gate and that this implies that $F$ contains a cycle as well, a contradiction showing that
$F^*$ is acyclic. The technical details of this derivation are provided in the lemma below. 
By construction, each wire connects output to input 
and there are no gates (except the input gates of course) having no input.
It follows that $F^*$ is a well formed circuit.

\begin{lemma} \label{nocycles}
$F^*$ has no cycles.
\end{lemma} 

{\bf Proof.} Observe first that if $F^*$ has a cycle involving at least two original gates $g_1$ and $g_2$
then we can conclude existence of such cycle in $F$, which will supply us a desired contradiction. 
Indeed, let $g_1, \dots, g_r$ be all the original gates of the cycle. Then, according to Lemma \ref{ubconnect} there is a connecting path between any two consecutive singleton gates and also between $g_r$ and $g_1$. Applying Lemmas \ref{andconnect}, and
\ref{notconnect} depending on the nature of the specific gates, we observe that in $F$ there are wires from each $g_i$ to $g_{i+1}$ (treating $r+1=1$) that is, $F$ has a cycle, a contradiction.

Furthermore, let us observe that existence of one original gate in a cycle implies existence of another one. Indeed, 
following the argumentation in the proof of Lemma \ref{ubconnect}, we observe that to arrive from a singleton gate to a singleton gate (even to itself) one has to go through an adjacency wire. However, the label on the other side of the adjacency wire is disjoint with the label of the tail side and thus when we start to descend through $in(S)$ gates we eventually (without closing the cycle before that since we have not arrived yet at the initial original gate!) will encounter another original gate, different from the starting one. Similar argumentation means that any $in$-gate in a cycle imply the presence of a singleton gate eventually. This rules out adjacency and parent-child arc from a potential cycle and leaves us only with child-parent arc but they are acyclic by construction since they go from a smaller label to a larger one. $\blacksquare$

In the rest of the discussion we implicitly assume that $F^*$ is well formed without explicit reference to Lemma \ref{nocycles}.

For each gate $g$ of $F^*$ denote by $f(g,F^*)$ the function
computed by a subcircuit of $F^*$ rooted by $g$. We establish properties of these functions from which
Theorem \ref{samefunctions} will follow by induction. 
In the following we sometimes refer to $f(g,F^*)$  as the function of $g$.

\begin{lemma} \label{notfunction}
For each {\sc not} gate $g$ of $F^*$, $f(g,F^*)$ is the negation of $f(g',F^*)$, where $g'$ is the input
of $g$ in $F$.
\end{lemma}

{\bf Proof.}
According to Lemma \ref{notconnect}, $F^*$ has a path from 
$g'$ to $g$ where all vertices except the first one are {\sc not} gates.
Since all of them but the last one are doubled, there is an odd number of such {\sc not} gates. Each {\sc not} gate has 
a single input, hence the function of each gate of the path (except the first one) is the negation of the function of its predecessor.
Hence these functions are, alternatively, the negation of the function of $g'$ and the function of $g'$. Since the number of 
{\sc not} gates in the path is odd, the function of $g$ is the negation of the function of $g'$, as required. $\blacksquare$

In order to establish a similar statement regarding {\sc and} and {\sc or} gates we need two auxiliary lemmas.

\begin{lemma} \label{childparentconnect}
For each label $S$, $f(oand(S),F^*)$ is the conjunction of
$f(g,F^*)$ of all original gates $g$ contained in $S$. 
Similarly, $f(oor(S),F^*)$ is the disjunction of the functions of such gates.
\end{lemma}

{\bf Proof.} 
We prove the lemma only for the $oand$ gates as for the $oor$ gates the proof is symmetric.
The proof easily goes by induction. For an original gate this is just a conjunction of a single element, namely itself, and 
this is clear by construction. For a larger label $S$, it follows by construction that $f(oand(S),F^*)=
f(oand(S_1),F^*) \wedge f(oand(S_2,F^*))$, where $S_1$ and $S_2$ are the children of $S$. For $S_1$ and $S_2$ the
rule holds by the induction assumption. Hence, $f(oand(S),F^*)$ is the conjunction of all the functions of all the original gates in the union of $S_1$ and $S_2$, the same as $f(oand(S),F^*)$ is the conjunction of the functions of all the original gates contained in $S$, as required.
$\blacksquare$

Let us call a path of $F^*$ \emph{semi-connecting} if it starts with an adjacency wire and the rest of the wires are parent-child ones.

\begin{lemma} \label{semiconnect}
Let $S$ be an {\sc and} label. Then $f(in(S),F^*)$ is the conjunction of the functions of all gates from which there is a semi-connecting path to $in(S)$.
For the {\sc or} label the statement is analogous with the conjunction replaced by disjunction.
\end{lemma}

{\bf Proof.} 
We provide the proof only for the {\sc and} label, for the {\sc or} label
the proof is analogous with the corresponding replacements of {\sc and} by {\sc or} and conjunctions by disjunctions.

The proof is by induction on the decreasing size 
of labels. For the largest {\sc and} label $S$, all the input wires are the adjacency wires. Clearly the considered function is the conjunction of the functions of the gates at the tails of these adjacency wires. It remains to see if there are no more gates to arrive at $in(S)$ by semi-connected paths. But any such gate, after passing through the adjacency wire must meet an ancestor of $S$ and, by the maximality assumption, $S$ has
no ancestors.

The same reasoning as above is valid for any label $S$ without ancestors. If $S$ has ancestors, then $f(in(S),F^*)$
is the conjunction of the functions of the gates at the tails
of the adjacency wires incident to $in(S)$ and the function of the $in$ gate of the parent of $S$ . By the induction assumption, this function is in fact a conjunction of the gates at the tails of the adjacency wires incident to $in(S)$ plus those connected to $in(S)$ by semi-connected paths through the parent. Since any semi-connected path either directly hits $in(S)$ at the head of an adjacency wire or approaches it through the parent, the statement is proven. $\blacksquare$


\begin{lemma} \label{andorfunction}
The function of any original {\sc and} gate $g$ of $F^*$
is the conjunction of the functions of the singleton gates whose outputs are the inputs of $g$ in $F$.
The same happens for the {\sc or} gate and the disjunction. 
\end{lemma}

{\bf Proof.}
As before, we prove the statement for the {\sc and} gate, for the {\sc or} gate it is analogous with the respective substitutions. 
By construction and Lemma \ref{semiconnect}, $f(g,F^*)$
is the conjunction of functions of all $oand$ gates (since there are no other ones) connected to $g$ by semi-connected paths. 
Let us call the labels of these $oand$ gates the \emph{critical labels}. Combining this with Lemma \ref{childparentconnect}, we see that $f(g,F^*)$ is in fact a conjunction of the functions
of all original gates contained in the critical labels. It remains to show that
these gates are exactly the in-neighbors of $g$ in $F$. Let us take a particular in-neighbor $g'$.
By Lemma \ref{andconnect}, there is a connecting path from $g'$ to $g$ and by Lemma \ref{childparentconnect},
the tail of the adjacency wire of this path is 
the $oand$ gate of a critical label, so $g'$ is in the required set. Conversely, assume that $g'$ is a gate in the required set.
Specify a critical label $S$ $g'$ belongs to. Clearly, there is a child-parent path from $g'$ to $oand(S)$ which, together with a semi-connected path from $oand(S)$ to $g$, makes a connecting path. The latter means that in $F$ there is a wire from $g'$ to $g$ according to Lemma \ref{andconnect}, as required.
$\blacksquare$


{\bf Proof of Theorem \ref{samefunctions}.}
Let us order the gates topologically and do induction on the topological order. The first gate is an input gate and the function of the input is just the corresponding variable both in $F$ and in $F^*$. Otherwise, the gate is {\sc and} or {\sc or} or {\sc not} gate. In the former two cases, according to Lemma \ref{andorfunction} the function of $g$ in $F^*$ is the conjunction (or disjunction, in case of {\sc or}) of the functions of its inputs in $F$, the same relation as in $F$. The theorem holds regarding the inputs by the induction assumption, hence the function of $g$ in $F^*$ is the same as in $F$. Regarding the {\sc not} gate, the argumentation is analogous, employing Lemma \ref{notfunction}. $\blacksquare$

\subsection{Proof of Theorem \ref{widths}} \label{twidth}

Let us define the undirected graph $H=H(T,{\bf G})$ called the
\emph{representation graph} of $(T,{\bf G})$ as follows. The vertices of this graph are the labels of $(T,{\bf G})$ 
and two vertices $S_1$ and $S_2$ are adjacent if and only if either $S_1$ is a child of $S_2$ (or vice versa of course) or $S_1$ and $S_2$ are adjacent in 
$(T,{\bf G})$ (meaning that the new adjacency operation is applied on $S_1$ and $S_2$).
We call the first type of edges \emph{child-parent} edges
and the second type \emph{adjacency} edges.

\begin{lemma} \label{widthf}
Let $t$ be the treewidth of $H$.
Then the treewidth of $F^*$ is at most $3t+2$.
\end{lemma}

{\bf Proof (Sketch).} Observe that if we contract the gates in $F^*$ of each label into a single vertex,
eliminate directions and remove multiple occurrences of edges, we obtain a graph isomorphic to $H$.
The desired tree decompositom is obtained from the tree decomposition of $H$ by replacing the occurrence of each vertex
of $H$ in a bag by the gates corresponding to this vertex. Thus, there is a tree decomposition of $F^*$
with at most $3(t+1)$ elements in each bag, that is the treewidth of $F^*$ is at most $3t+2$. $\blacksquare$

\begin{lemma} \label{widthh}
The treewidth of $H$ is at most $k$,
where $k$ is the width of $(T,{\bf G})$. 
\end{lemma}

{\bf Proof.}
For each node $t$ of $T$, let $S(t)$ be the set of labels of the graph
associated with $t$. Consider the structure $(T,{\bf B})$ where
${\bf B}$ is a family of subsets of $H$ associating with each node
$t$ a set $B(t)$ consisting of vertices of $H$ corresponding to the 
elements of $S(t)$. We are going to show that $(T,{\bf B})$ is a tree
decomposition of graph $H'$ obtained from $H$ by removal of all child-parent
edges.

First of all, observe that for each $v \in V(H)$, the subgraph $T_v$ of $T$
consisting of all nodes containing $v$ is a subtree of $T$. 
Let us consider $T$ as a rooted tree
with the root $t$ being the same as in $(T,{\bf G})$.
Let $t_1$ and $t_2$ be two nodes containing $v$. Then one of them
is an ancestor of the other. Indeed, otherwise $t_1$ and $t_2$ are nodes of two disjoint
subtrees $T_1$ and $T_2$ whose roots $t'_1$ and $t'_2$ are children of some node $t^*$.
By the definition of {\sc scd}, $G(t'_1)$ is disjoint with $G(t'_2)$ and it is
not hard to conclude from the definition that $V(G(t_1)) \subseteq V(G(t'_1))$ and 
$V(G(t_2)) \subseteq V(G(t'_2))$ are disjoint. Since any label is a subset of the set of vertices of 
the graph it belongs to, $S(t_1)$ and $S(t_2)$ cannot have a common label and hence
$B(t_1)$ and $B(t_2)$ cannot have a joint node. 
Furthermore, it is not hard to observe, if $t_1$ is ancestor of $t_2$ and $S \in S(t_1) \cap S(t_2)$ 
then $S$ belongs to $S(t')$ of all nodes $t'$ in the path between $t_1$ and $t_2$. Of course, the same
is true regarding the node of $H$ corresponding to $S$.
Thus we have shown that if $t_1$ and $t_2$ contain $v$ they cannot belong to different 
connected components of $T_v$, confirming the connectedness of $T_v$.

Next, we observe that if $v_1$ and $v_2$ are incident to an adjacency edge then
there is a node $t$ containing both $v_1$ and $v_2$. Indeed, let $S_1$ and $S_2$
be the labels corresponding to $v_1$ and $v_2$, respectively. Let $t$ be the node
where the adjacency operation regarding $S_1$ and $S_2$ is applied. Then both $S_1$
and $S_2$ belong to $S(t)$ and, consequently, $t$ contains both $v_1$ and $v_2$.
Finally, by construction, each vertex of $H$ is contained in some node.

To obtain the desired tree decomposition of $H$, we are going to modify $(T,{\bf B})$
to acquire two properties: that the number of nodes of the resulting tree is at most $2|F|$
and that each parent-child pair $u,v$ is contained in some node $t$. 
For the former just iteratively remove all nodes whose operations are new adjacency.
If the node $t$ being removed is not the root then make the parent of $t$ to be the parent
of the only child of $t$ (since $t$ has only one child the tree remains binary).
The latter property can be
established by adding at most one vertex to each bag of the resulting structure $(T',{\bf B'})$. Indeed,
for each non-singleton label $S$, let $t(S)$ be the node where this label is created by the union 
operation. Then both children of $S$ belong to the only child of $t(S)$.
Let $(T',{\bf B^*})$ be obtained from $(T',{\bf B'})$ as follows. For each non-singleton label
$S$, add the vertex corresponding to $S$ to the bag of the child of $t(S)$. Since at most one new label
is created per node of $T'$, at most one vertex is added to each bag. It is not hard to see 
both of the modifications preserve
properties stated in the previous paragraphs and achieve the desired properties regarding the child-parent edges.
Since each bag of $(T,{\bf B'})$ contains at most $k+1$ elements, we conclude that the treewidth of $H$
is at most $k$. Since the number of bags is at most as the number of labels, we conclude that the number of bags
is at most $2|F|$ 
$\blacksquare$

{\bf Proof of Theorem \ref{widths}.}
Immediately follows from the combination of
Lemmas \ref{widthf} and Lemma \ref{widthh}. 
$\blacksquare$

\section{Application to knowledge compilation}
In this section we demonstrate an application of Theorem \ref{finalwidth} to knowledge compilation
by showing existence of an algorithm compiling the given circuit $Z$ into {\sc dnnf}. Both the time
complexity of the algorithm and the space complexity of the resulting {\sc dnnf} are fixed-parameter linear
parameterized by the cliquewidth of $Z$. More precisely, the statement is the following:

\begin{theorem} \label{mainone}
Given a single-output circuit $Z$ of cliquewidth $k$,
there is a {\sc dnnf} of $Z$  having size $O(9^{18}k^2|Z|)$.
Moreover, given a clique decomposition of $Z$ of width $k$, there is a 
$O(9^{18k}k^2|Z|)$ algorithm constructing such a {\sc dnnf}.
\end{theorem}

Theorem \ref{mainone} is an immediate corollary of Theorem \ref{finalwidth} and the following one:
\begin{theorem} \label{dnnfcircuit}
Given a single-output circuit $Z$ of treewidth $p$, there is  a {\sc dnnf} of $Z$ having size $O(9^pp^2|Z|)$.
Moreover, such a {\sc dnnf} can be constructed by an algorithm of the same runtime that gets
as input the circuit $Z$ and a tree decomposition of $Z$ of width $p$ having $O(Z)$ bags.
\end{theorem}

The rest of this section is a proof of Theorem \ref{dnnfcircuit}.
Our first step is Tseitin transformation from circuit $Z$ into a {\sc cnf} $F'$.
For this purpose we assume that $Z$ does not have paths of $2$ or more {\sc not} gates.
Depending on whether this path is of odd or even length, it can be replaced by a single
{\sc not} gate or by a wire, without treewidth increase. In this case the variables $y_1, \dots, y_m$
of $F'$ are the variables of $Z$ and the outputs of {\sc and} and {\sc or} gates of $Z$.
Under this assumption, it is not hard to see that the inputs of each gate are literals 
of $y_1, \dots, y_m$. Then the output $x$ of $Z$ is either $y_i$ or $\neg y_i$ for some $i$.
Let us call $x$ the output literal. 

The {\sc cnf} $F'$ is a conjunction of the singleton clause containing the output literal and the 
{\sc cnf}s associated with each {\sc and} and {\sc or} gate. Let $C$ be an {\sc and} gate with inputs 
$t_1, \dots, t_r$ and output $z$. Then the resulting {\sc cnf} is 
$(t_1 \vee \neg z) \wedge \dots \wedge(t_r \vee \neg z) \wedge (\neg t_1 \vee \dots \vee \neg t_r \vee z)$. 
If $C$ is an {\sc or} gate then the resulting {\sc cnf} is 
$(\neg t_1 \vee z) \wedge \dots \wedge (\neg t_r \vee z) \wedge (t_1 \vee \dots \vee t_r \vee \neg z)$.
We call the last clause of the {\sc cnf} of $C$ the \emph{carrying} clause w.r.t. $C$ and the rest 
are \emph{auxiliary} ones w.r.t. $C$ and the corresponding input. 

To formulate the property of Tseitin transformation that we need for our transformation, let us extend
the notation. We consider sets of literals that do not contain a variable and its negation.
For a set $S$ of literals, $Var(S)$ is the set of variables of $S$. 
The \emph{projection} $Pr(S,V')$ of $S$ to a set $V'$ of variables is the subset $S'$ of $S$ 
obtained by the removal of variables that are not in $V'$.
Let ${\bf S}$ be a family of sets of literals over a set $V$ of variables. Then the projection
$Pr({\bf S},V')$ of ${\bf S}$ to $V' \subseteq V$ is $\{Pr(S,V')|S \in {\bf S}\}$.
Denote by $Var(Z)$ and $Var(F')$ the sets of variables of $Z$ and $F'$, respectively. Let us say
that a set $S$ of literals with $Var(S)=Var(Z)$ is a \emph{satisfying assignment} of $Z$ if $Z$ is
true on the truth assignment on $Var(Z)$ that assigns all the literals of $S$ to true. For a {\sc cnf},
the definition is analogous. The well known property of Tseitin transformation is the following:

\begin{lemma} \label{tseitin}
Let ${\bf S_1}$ and ${\bf S_2}$ be the sets of satisfying assignments of $F'$ and $Z$,
respectively. Then $Pr({\bf S_1},Var(Z))={\bf S_2}$.
\end{lemma}

Lemma \ref{tseitin} is useful because of the following nice property of {\sc dnnf}.
\begin{lemma} \label{dnnfproj} (Theorem 9 of \cite{DarwicheJACM}).
Let $Z$ be a DNNF let $V' \subseteq Var(Z)$ and let $Z'$ be the {\sc dnnf} obtained from $Z$
by replacing the variables of $Var(Z) \setminus V'$ with the $true$ constant.
Let ${\bf S}$ and ${\bf S'}$ be sets of satisfying assignments of $Z$ and $Z'$, respectively.
Then ${\bf S'}=Pr({\bf S},V')$. 
\end{lemma}

Thus it follows from Lemmas \ref{tseitin} and \ref{dnnfproj} that having compiled $F'$ into a {\sc dnnf} $D'$,
a {\sc dnnf} $D$ of $Z$ can be obtained by replacing the variables of $Var(F') \setminus Var(Z)$ with the $true$
constant. Clearly, this does not incur any additional gates. In order to obtain a {\sc dnnf} of $F'$, we observe 
that the treewidth of the incidence graph of $F'$ is not much larger than the treewidth of $Z$.

\begin{lemma} \label{tseitindecomp}
Let $(T,{\bf B})$ be a tree decompositoion of $Z$ of width $p$.
There is a $O(p^2|T|)$ time algorithm ($|T|$ is the number of nodes of $T$) transforming
$(T,{\bf B})$ into a tree decomposition $(T^*,{\bf B^*})$ of the incidence graph $G'$ of 
$F'$ having width at most $2p+1$ and with $|T^*|=O(p^2|T|)$.
\end{lemma}

{\bf Proof.}
Let $F''$ be the {\sc cnf} obtained from $F'$ by removal of all the clauses but the carrying ones and let $G''$
be the respective incidence graph. Transform $(T,{\bf B})$ into $(T,{\bf B''})$ as follows:
\begin{itemize}
\item Replace each occurrence of an {\sc and} or {\sc or} gate $X$ with the respective carrying clause and the
      variable corresponding to the output of $X$.
\item Replace each occurrence of a {\sc not} gate with the variable corresponding to the input of the gate
      (it may either be an input variable of $Z$ or the output variable of some {\sc and} or {\sc or} gate).			
\end{itemize}
Let us show that $(T,{\bf B''})$ is indeed a tree decomposition 
of $G''$ of width $2p+1$.

Each element of a bag of ${\bf B}$ is replaced by at most $2$ elements, hence the size of a bag
is at most twice the maximal size of bag of ${\bf B}$, i.e. at most $2(p+1)$. Consequently the
width of $(T,{\bf B''})$ is at most $2p+1$. Let us verify the connectedness property.
An original variable $x$ of $Z$ is contained in a node $t$ of $(T,{\bf B''})$ if and only if
in $(T,{\bf B})$ $t$ contains either $x$ or the NOT gate $Y$ with input $x$. By the connectedness property
both nodes of $(T,{\bf B})$ containing $x$ and those containing $Y$ form subtrees and by the adjacency
property, these subtrees have at least one joint vertex. It follows that their union forms a subtree of $T$.
Each new variable $y$ corresponding to a gate $C$ of $Z$ is contained in exactly those nodes of $(T,{\bf B''})$
that contain $C$ or the negation of $C$ in $(T,{\bf B})$. Exactly the same argument as in the previous case
ensures connectedness regarding $y$. Finally each carrying clause $C$ is contained in exactly those nodes of $(T,{\bf B''})$
that contain the corresponding gate in $(T,{\bf B})$. So, the connectedness regarding $C$ follows from the connectedness
property of $(T,{\bf B})$. Thus we have established the connectedness of $(T,{\bf B''})$.
To establish the adjacency property, let $C$ be a carrying clause corresponding to a gate $X$ in $C$ and 
let $v$ be a variable occurring in $C$. If $v$ corresponds to the output of $C$ then the adjacency follows by
construction because $v$ is explicitly put in those clauses where $X$ appears. So, assume that $v$ corresponds to
an input of $X$. If $v$ is an original variable then $(T,{\bf B})$ has a node $t$ containing a literal of $v$ and $X$.
By construction, in $(T,{\bf B''})$, $t$ contains $v$ and $C$. So, assume that $v$ is the output variable of some gate
$X'$ and let $C'$ be the corresponding carrying clause of $X'$. It follows that in $(T,{\bf B})$ there is anode $t$
containing both $X$ and $X'$. Consequently, in $(T,{\bf B''})$ the same node $t$ contains $C,C',v$. So, the adjacency 
property has been established and we conclude that $(T,{\bf B''})$ is indeed a tree decomposition of $G''$ of
width $2p+1$.

Next, we observe that for each {\sc and} or {\sc or} gate $X$ of $Z$ and for each variable $u$ of $F'$ corresponding to an input
of $X$ and for variable $y$ of $F'$ corresponding to the output of $X$, there is a node $t$ of $(T,{\bf B''})$
containing both $y$ and $u$. Indeed, let $C$ be the carrying clause corresponding to $X$. By construction, whenever
$t$ contains $C$, $t$ also contains $y$. By the adjacency property, there is at least one $t$ containing $C$ and $u$.
Since this last $t$ contains also $y$, this is a desired clause. Pick one node with the specified property and denote
it by $t(y,u)$. Add to $T$ a new node $t'$ with $t(y,u)$ being its only neighbor. The bag of $t'$ will contain
$y,u$, and $C(y,u)$ the auxiliary clause of $X$ corresponding to the input $u$. Do so for all the auxiliary clauses. 
Finally, let $y$ be the variable occurring in the output clause (the clause containing the 
output literal). Specify a node $t$ containing this variable.
Add a new node $t'$ for whom $t$ is the only neighbor and add $y$ and the output clause to the bag of $t'$. 
Let $(T^*,{\bf B^*})$ be the resulting structure. Clearly the connectedness is preserved and the adjacency property
is established for the clauses of $F'$ that are not included in $F''$. It follows that $(T^*,{\bf B^*})$ is a tree
decomposition of $G'$ by construction, its width does not exceed the width of $(T,{\bf B'})$, i.e. at most
$2p+1$ and the additional $O(p^2|Z|)$ nodes (their number is bounded by the number of wires of $Z$ plus $1$ for
the output clause) are leaves. The desired runtime of the transformation from $(T,{\bf B})$ to
$(T^*,{\bf B^*})$ clearly follows from the above description. 
$\blacksquare$

It remains to show that a space-efficient {\sc dnnf} can be created parameterized by the treewidth of the incidence
graph. 

\begin{theorem} \label{dnnfcnf}
Let $F$ be a {\sc cnf} and let $(T',{\bf B'})$ be a tree decomposition of the incidence graph of $F$.
Then $F$ has a {\sc dnnf} of size $O(3^t|T'|)$ where $t$ is the width of $(T',{\bf B'})$.
Moreover, given $F$ and $(T',{\bf B'})$ such a {\sc dnnf} can be constructed by an algorithm having the same runtime. 
\end{theorem}

The proof of Theorem \ref{dnnfcnf} is provided in Section \ref{dnnfcnfproof}.

{\bf Proof of Theorem \ref{dnnfcircuit}.}
The construction of a {\sc dnnf} for $Z$ consists of $4$ stages: transform $Z$ into $F'$ by the Tseitin transformation;
transform the tree decomposition of $Z$ into a tree decomposition of the incidence graph of $F'$; obtain a {\sc dnnf}
of $F'$ as specified by Theorem \ref{dnnfcnf} and obtain a {\sc dnnf} of $Z$ as specified in Lemma \ref{dnnfproj}.
The correctness of this procedure follows from the above discussion. The time and space complexities easily follow from 
the combination of the complexities of intermediate stages. $\blacksquare$

\subsection{Proof of Theorem \ref{dnnfcnf}} \label{dnnfcnfproof}

The proof of Theorem \ref{dnnfcnf} is based on the same idea as the proof that a CNF with the width of the primary graph
at most $p$ has a {\sc dnnf} of size $O(2^pn)$ \cite{DarwicheJACM}. The difference is that we have to take into account that
the bags of the tree decomposition contain clauses as well as variables. Let us introduce notation. Let $F$ be the CNF whose
{\sc dnnf} we are going to construct,
$G$ be the incidence graph of $F$ and $(T,{\bf B})$ be a tree decomposition of $G$. In what follows we identify the vertices
of $G$ with the respective variables and clauses. For each node $t$ of $T$, we denote the bag of $T$ by $B(t)$. 
Recall that for an element $a \in B(t)$ (either a variable or clause), we say 
that $t$ \emph{contains} $a$. 
We assume that $(T,{\bf B})$ is a \emph{minimal} tree decomposition in the sense that removal of
any element from a bag violates a tree decomposition property. This assumption is not constraining
because such tree decomposition is easy to obtain by iterative removal of nodes from the bags until no
further removal is possible.

We pick an arbitrary node $tr$ of $T$ and let to be the root and in what follows
we consider $T$ to be a rooted tree. We assume w.l.o.g. that $T$ has at most $2$ children. 
Indeed, otherwise, if some node $t$ has
children $t_1, \dots, t_r$ for $r>2$, we introduce additional nodes $t'_2, \dots, t'_r$ make the sequence $t,t'_2, \dots, t'_r$ going from the parent to a child, $t_1$ remains a child of $t$ and for each $t'_i$, node $t_i$ becomes the additional child.
The bags of $t'_2, \dots, t'_r$ are made identical to the bag of $t$. Such transformation increases the number of nodes at most twice and hence proving the theorem for such transformed tree preserves the desired asymptotic. 

We consider only sets of literals with at most one literal per variable. For a set $S$ of literals, let $Var(S)$
be the set of variables whose literals occur in $S$.
We denote by $Cl(F)$ and $Var(F)$ the set of clauses and variables of $F$. For a node $t$ of $T$ we denote by $Cl(t)$
and $Var(t)$ the set of clauses and variables contained in $t$. For a subtree $T'$ of $T$, $Cl(T')$ and $Var(T')$ denote
the set of clauses and variables contained in the nodes of $T'$. For a clause $C$ and a set $V$ of variables, we denote
by $Pr(C,V)$ the \emph{projection} of $C$ to $V$ i.e. the clause obtained by the removal from $C$ the occurrences
of all the variables that are not in $V$. Recall that for a set $S$ of literals, we use $Pr(S,V)$ with the analogous meaning. 
For a CNF $F'$, we denote by $Pr(F',V)$ the CNF obtained from $F'$ by
projecting all of its clauses to $V$. 
For a subtree $T'$ of $T$, we denote $Pr(F,Var(T'))$ by $F(T')$.

Let us call two circuits (formulas including CNF are regarded a special cases of circuits)
equivalent if they have the same set of variables and the same set of satisfying assignments.
One way to create a formula equivalent to the given CNF is \emph{Shannon expansion}. Let $F$
be a CNF and let $x$ be a variable of $F$. Then $F|x$ denotes the CNF obtained from $F$ by removal
of all the clauses containing $x$ and removal from of all the occurrences of $\neg x$ from the 
remaining clauses. It is known that $(F|x)x \vee (F| \neg x) \neg x$ is equivalent to $F$. Applying this 
transformation over a set $V$ of variables works as follows. Let $S$ be a set of literals such
that $Var(S)=V$. Analogously to $F|x$, $F|S$ is the CNF obtained from $F$ by removal of all the
clauses containing the occurrences of $S$ and removal of the occurrences of the opposite literals
from the remaining clauses. Let us call the disjunction $\bigvee_{Var(S)=V}(F|S)S$ the
\emph{generalized Shannon expansion} of $F$ w.r.t. $V$. Applying the Shannon expansion inductively, it is
not hard to show that the generalized Shannon expansion of $F$ w.r.t. $V$ is equivalent to $F$.

Let us extend our notation. We denote by $F \setminus C'$ the set of clauses obtained from
$F$ by removal of all the clauses of $C'$. Let $t'$ be the root of $T'$ and let $C' \subseteq Cl(t')$ and let
$S$ be a set of literals assigning a set of variables $V' \subseteq Var(t')$.
We denote $(Pr(F \setminus C',Var(T')))|S$ by $F(T',C',S)$ and call it a \emph{residual} of 
$F(T')$ (induced by $C'$ and $S$ if the context requires mentioning it).
When $S$ or $C'$ is empty, we can use $F(T',C')$ and $F(T',S)$ with the obvious meaning.
If $S$ assigns all the variables contained in $t'$, we say that $F(T',C',S)$ is a \emph{basic residual} (BR).
Finally, we call the function $(\bigwedge S') \wedge F(T',C',S)$ \emph{extended residual} (ER) ($\bigwedge S'$
means conjunction of all the literals of $S' \subseteq S$) or, in case $Var(S)=Var(t')$, \emph{extended basic residual} (EBR).

\begin{lemma} \label{localdisj}
Any residual or extended residual of $F(T')$ is equivalent to a disjunction of EBRs of $F(T')$. 
\end{lemma}
 
{\bf Proof.}
Let $F(T',C',S)$ be a residual of $F(T')$.
Let $F_1$ be a disjunction over all $F(T',C',S \cup S_1) \wedge \bigwedge S_1$ where $S_1$ is a set
of literals of $Var(t') \setminus Var(S)$. Then $F(T',C',S)$ is equivalent to $F_1$ because $F_1$
is a generalized Shannon expansion of $F(T',C',S)$ w.r.t $Var(S_1)$ 
$\blacksquare$

\begin{lemma} \label{unnode}
Let $t_1$ be a child of $t'$ and let $T_1$ be the subtree rooted by $t_1$.
Let $C$ be a clause of $F$ containing an occurrence of a variable $x \in Var(T_1) \setminus Var(t')$.
Then $C$ is contained in a node of $T_1$.
\end{lemma}

{\bf Proof.}
By the adjacency property, there must be a node $t''$ of $T$ containing both $C$ and $x$.
This node cannot be $t'$ by definition. This node cannot be anyone outside $T_1$ because
otherwise, by the connectedness property, it will be required that $x$ is contained in
$t'$ in contradiction to our assumption. It remains to conclude that $t''$ is a node
of $T_1$. $\blacksquare$

\begin{lemma} \label{unclause}
Let $t_1$ and $t_2$ be $2$ children of $T'$ and let $T_1$ and $T_2$ be the
subtrees of $T$ rooted by them. Let $C \in Cl(t_1) \cap Cl(t_2)$.
Then $C$ contains occurrences of $Var(T_1) \setminus Var(t')$ and
of $Var(T_2) \setminus Var(t')$.
\end{lemma}

{\bf Proof.}
Assume that $C$ does not contain occurrences of, say, $Var(T_1) \setminus Var(t')$.
We claim that the occurrences of $C$ can be removed from all the nodes of $T_1$
in contradiction to the minimality of $(T,{\bf B})$. This removal clearly does not
violate the connectedness property because the path between any two nodes outside
of $T_1$ does not go through $T_1$. As for adjacency property, let $x$ be any variable
contained together with $C$ in a node of $T_1$. 
If $C$ and $x$ are adjacent then $x \in Var(T_1) \cap Var(t')$ and therefore,
their adjacency is witnessed by the bag of $t'$. $\blacksquare$

Let $F(T',C',S)$ be a basic residual of $F(T')$.
Let $C'' \subseteq Cl(t')$ be the union of $C'$ and the set of clauses of $t'$ satisfied 
by $S$. We call the set $Pr((Cl(t_1) \cap Cl(t_2)) \setminus C'',Var(T'))|S$ the \emph{branching set}
of $F(T',C',S)$.

\begin{lemma} \label{t1t2}
Let $t_1$ and $t_2$ be the children of $t'$ rooting repective subtrees $T_1$ and $T_2$.
Then $Var(F(T',C',S)) \subseteq Var(T_1) \Delta Var(T_2)$. Moreover,the set of clauses
containing occurrences of both $Var(T_1) \setminus Var(T_2)$ and $Var(T_2)\setminus Var(T_1)$
is precisely the branching set of $F(T',C',S)$. 
\end{lemma}

{\bf Proof.}
By construction, $Var(F(T',C',S)) \subseteq Var(T')$. Furthermore, $Var(T')=(Var(T_1) \Delta Var(T_2)) \cup Var(t')$.
Since $S$ assigns $Var(t')$, it follows that\\ $Var(F(T',C',S)) \subseteq Var(T_1) \Delta Var(T_2)$.

For the second statement, let $C$ be a clause of $F(T',C',S)$ containing entries of both
$Var(T_1) \setminus Var(T_2)$ and $Var(T_2) \setminus Var(T_1)$. This means that there is a clause $C^{or}$ of $F$
such that $C=Pr(C^{or},Var(T'))|S$. According to Lemma \ref{unnode}, $C^{or} \in Cl(t_1) \cap Cl(t_2)$.
By construction, $C^{or} \notin C''$. It follows that $Pr((Cl(t_1) \cap Cl(t_2)) \setminus C'',Var(T'))|S$ in particular contains
$Pr(C^{or},Var(T'))|S=C$. 

Conversely, assume that $C$ belongs to the branching set of $F(T',C',S)$.
It follows that there is a clause $C^{or} \in Cl(t_1) \cap Cl(t_2) \setminus C''$ related to $C$ as defined above.
By the connectedness property, $Cl(t_1) \cap Cl(t_2) \subseteq Cl(t')$ that is, $C^{or}$ is contained in $t'$.
Consequently, since $C^{or} \notin C''$, we conclude that $C^{or}$ is not satisfied by $S$.
It follows that $C=Pr(C^{or},Var(T'))|S$ is a clause of $F(T',C',S)$.
According to Lemma \ref{unclause}, $C^{or}$ contains occurrences of $x \in Var(T_1) \setminus Var(t')$ and
$y \in Var(T_2) \setminus Var(t')$. By definition, both $x$ and $y$ belong to $Var(T') \setminus Var(S)$ and
hence they are preserved in $C$. By the connectednes property, $x \in Var(T_1) \setminus Var(T_2)$ and
$y \in Var(T_2) \setminus Var(T_1)$, hence the opposite direction holds. $\blacksquare$   

Another method of equivalence preserving transformation is \emph{clausal expansion}. Let $C$ be a clause
of a CNF $F$ and $I_1,I_2$ be a partition of $C$. Then, it follows from De Morgan laws that
$(F \setminus \{C\} \wedge I_1) \vee (F \setminus \{C\} \wedge I_2)$ is equivalent to $F$.
We extend this to the generalized clausal expansion. Let $C^*$ be a set of clauses of $F$.
For each $C \in C^*$, define a partition $I_1(C),I_2(C)$. Let ${\bf I}$ be the set of all CNFs
$I$ whose set of clauses are exactly one of $I_1(C),I_2(C)$ for each $C \in C^*$.
Then $\bigvee_{I \in {\bf I}} (F \setminus C^*) \wedge I$ is equivalent to $F$ and called a
\emph{generalized clausal expansion} of $F$ w.r.t. ${\bf I}$. 

\begin{lemma} \label{globaldisj}
Let $T'$ be a subtree of $T$ with root $t'$ and assume that $t'$ has two children $t_1$ and $t_2$ and let $T_1$
and $T_2$ be the subtrees of $T$ rooted by $t_1$ and $t_2$, respectively. Then each basic residual $F(T',C',S)$
of $F(T',C',S)$ is either  unsatisfiable or can be represented (for some $r$) 
as a $(F_1 \vee \dots \vee F_r)$ where 
each $F_i$ is a conjunction of a residual of $F(T_1)$ and a residual of $F(T_2)$.
Moreover the number of such conjunctions of residuals that are needed to represent \emph{all} the basic residuals
$F(T',C',S)$ does not exceed $3^p$, where $p$ is the width of $(T,{\bf B})$. 
\end{lemma}

{\bf Proof.}
The unsatisfiability clearly follows if $F(T',C',S)$ contains an empty clause.
Otherwise, according to Lemma \ref{t1t2}, all the occurrences of each clause of
$F(T',C',S)$ belong to $Var(T_1)\Delta Var(T_2)$. 
Let $C^*$ be the set of clauses of $F(T',C',S)$ containing occurrences
of variables of both $Var(T_1) \setminus Var(T_2)$ and of $Var(T_2) \setminus Var(T_1)$.
For each $C \in C^*$, let $I_1(C),I_2(C)$ be the partition of $C$ into literals of variables
of $Var(T_1) \setminus Var(T_2)$ and of $Var(T_2) \setminus Var(T_1)$.
Let ${\bf I}$ be the set of CNFs obtained from $C^*$ by taking exactly one of $I_1(C)$ or $I_2(C)$
for each $C \in C^*$. Let $F^*$ be the clausal expansion of $F(T',C',S)$ w.r.t. $C^*$ and ${\bf I}$.
We know that $F^*$ is equivalent to $F(T',C',S)$. 

By construction, $F^*$ is a disjunction of CNFs.
Let $F''$ be one of the disjuncts. Again by construction, $F''$ can be reprsented as the conjunction
of CNFs $F'_1$ whose clauses contain occurrences of $Var(T_1) \setminus Var(T_2)$ only and $F'_2$,
whose clauses contain occurrences of $Var(T_2) \setminus Var(T_1)$. Let $I_1$ be the subset of 
$C^*$ consisting of the clauses $C \in C^*$ such that $I_2(C)$ is taken to $F''$ by the clausal
expansion and let $I_2=C^* \setminus I_1$, i.e. those clauses $C$ of $C^*$ that $I_1(C)$ is taken
to $F''$. According  to Lemma \ref{t1t2}, $C^*$ is nothing else than the branching set of $F(T',C',S)$.
It follows that for each $C \in C^*$ there is a set $CC^{or} \subseteq (Cl(t_1) \cap Cl(t_2)) \setminus C''$
such that $\{C\}=Pr(CC^{or},Var(T'))|S$. Let $I^*_1=\bigcup_{C \in I_1} CC^{or}$ and let 
$I^*_2=\bigcup_{C \in I_2} CC^{or}$. Further on, let $S_1=Pr(S,Var(T_1))$ and let $S_2=Pr(S,Var(T_2))$.
Let $F_1=F(T_1,(C'' \cap Cl(t_1)) \cup I^*_1,S_1)$ and let $F_2=F(T_2,(C'' \cap Cl(t_2)) \cup I^*_2,S_2)$.
We claim that $F'_1=F_1$ and $F'_2=F_2$. We will prove only the former for the latter is symmetric.

Let $C$ be a clause of $F'_1$. This means that there is a clause $C^{or}$ of $F$ and the clause
$C^{int}$ of $F(T',C',S)$ such that $C^{int}=Pr(C^{or},Var(T'))|S$ and $C=C^{int}$ if $C^{int} \notin C^*$ or
$C=Pr(C^{int},Var(T_1))$ otherwise. In any case $C=Pr(Pr(C^{or},Var(T'))|S,Var(T_1))$.
Since $C^{or}$ is not satisfied by $S$, the operation $|S$ applied to any its subset is in
fact the projection to $Var(T)\setminus Var(S)$. With this in mind, we can write
$$C=Pr(Pr(C^{or},Var(T')),Var(T_1)|S=Pr(C^{or},Var(T') \cap Var(T_1))|S=Pr(C^{or},Var(T_1))|S$$
According to our assumption, $C$ is not empty. Let $x$ be a variable occurring in $C$.
By construction, $x \in Var(T_1)\setminus Var(t')$. It follows from Lemma \ref{unnode}
that $C^{or}$ is contained in a node of $Var(T_1)$. By definition, $C^{or}$ is not satisfied
by $S$, and does not belong to $C'$, from whence it follows that $C \notin C''$.  
Also, by definition, if $C^{or}$ belongs to a branching set then $C^{or} \in I^*_2$.
It follows that $C^{or} \notin C'' \cup I^*_1$ and that $C^{or}$ is not satisfied by $S_1$.
Together with the fact that $C^{or}$ is contained in a node of $T_1$, this implies that
$Pr(C^{or},Var(T_1))|S=C$ is a clause of $F_1$.

Conversely, let $C$ be a clause of $F_1$. Then there is a clause $C^{or}$ of $F$ that is contained
in some node of $T_1$ such that $C=Pr(C^{or},Var(T_1))|S_1$. Then $C^{or}$ is not satisfied by $S$.
Indeed, $C^{or}$ is not satisfied by $S_1$ by definition of $F_1$. If $C^{or}$ is satisfied by
any element of $S \setminus S_1$ this means that $C$ is contained in a node outside $T_1$ and hence,
by the connectedness property, $C$ is contained in $t'$. But then $C \in C''$ in contradiction to the 
definition of $F_1$. Taking into account that $C' \subseteq C''$, $C^{int}=Pr(C^{or},Var(T'))|S$
is a clause of $F(T',C',S)$.  If $C^{int} \in C^*$ then, by definition of $F_1$, $C^{or} \in I^*_2$
and hence $Pr(C^{int},Var_1)=C$ is a clause of $F'_1$. If $C^{int} \notin C^*$ then, according to
Lemma \ref{t1t2}, all the occurrences of $C^{int}$ are of $Var(T_1) \setminus Var(T_2)$ or all the
occurrences of $C^{int}$ are of $Var(T_2) \setminus Var(T_1)$. Notice that the latter case causes
contradiction. Indeed, since $F(T',C',S)$ does not contain empty clauses, $C^{int}$ in particular is
not empty. Let $x$ be a variable occurring in $C^{int}$. By our assumption $x \in Var(T_2)$. 
In fact, by construction of $F(T',C',S)$, $x \in Var(T_2) \setminus Var(t')$. It follows from Lemma \ref{unnode}
that $C^{or}$ is contained in a node of $T_2$. Since $C^{or}$ is also contained in a node of $T_1$, it follows from the
connectedness property that $C^{or} \in Cl(t_1) \cap Cl(t_2)$. Furthermore, $C^{or} \notin C''$ by definition
of $F_1$. Consequently, $C^{int}$ belongs to the branching set of $F(T',C',S)$. By Lemma \ref{t1t2},
$C^{int} \in C^*$ in contradiction to our assumption.It remains to assume that all the entries of $C^{int}$
belong to $Var(T_1) \setminus Var(T_2)$. In this case $C^{int}=Pr(C^{int},Var(T_1))=C$ is a clause of $F'_1$
as required.

Now, let us calculate the number of conjunctions $F_1 \wedge F_2$ needed to represent all the EBRs of $T'$.
Each $F_1 \wedge F_2$ is unambiguously determined by the set
$C''$, the partition $I^*_1,I^*_2$ of $Cl(t_1) \cap Cl(t_2) \setminus C''$ and the assignment of variables
contained in $t'$ 
Let $p_1,p_2$ be the number of clauses and variables contained in $t'$, respectively. It follows that there
are at most $3^{p_1}2^{p_2} \leq 3^{p_1+p_2} \leq 3^p$ choices for $F_1 \wedge F_2$, as required. 
$\blacksquare$


\begin{lemma} \label{layerdnnf}
Let $t',T',t_1,T_1,t_2,T_2$ be as in the statement of Lemma \ref{globaldisj} and let $D$ be a DNNF  
containing gates with outputs computing all the basic residuals of $F(T_1)$ and of $F(T_2)$.
Then a DNNF computing all the basic residuals of $F(T')$ can be computed by adding $O(3^p)$ new
gates, where $p$ is the width of $(T,{\bf B})$.
\end{lemma}

{\bf Proof.}
Each basic residual $F(T',C',S)$ is unambiguously defined by the set of clauses $C'$
and  the assignment $S$ to the variables contained in $t'$. Let $p_1$ be the number of clauses
and $p_2$ be the number of variables. Then the number of choices is at most $2^{p_1}*2^{p_2}=2^p$.
The DNNF being constructed will have at most $2^p$ {\sc or} gates whose outputs are the BRs of $F(T')$ 
and the inputs are conjunctions of residuals of $F(T_1)$ and $F(T_2)$ as specified in
Lemma \ref{globaldisj}.  
Let $F_1 \wedge F_2$ be a conjunction of a residual of $F(T_1)$ and a residual of $F(T_2)$. The number 
of such conjunctions needed to form inputs of the above {\sc or} gates is at most $3^p$ according to 
Lemma \ref{globaldisj}. Each $F_i$ is formed as a disjunction of EBRs of $F(T_i)$, according to Lemma 
\ref{localdisj}, contributing another $2*3^p$ to the overall number of gates.
 
Let us calculate the number of EBRs of $F(T_1)$ we need in order to compute
all the required residuals of $F(T_1)$. For $F(T_2)$ the calculation will be analogous.
For each residual $F(T_1,C_1,S_1)$ participating in a conjunction $F_1 \wedge F_2$ as above, 
$Var(S_1)=Var(t_1)  \cap Var(t')$ (see the construction
in the proof of Lemma \ref{globaldisj}). Let $p_1=|Var(S_1)|$ and $p_2=|Cl(t_1)|$.
Then the number of ways to form the residual is at most $2^{p_1+p_2}$. 
Applying the generalized Shannon expansion, we observe that each EBR
participating in the disjunction forming $F(T_1,C_1,S_1)$ is in the form
$\bigwedge S_3 \wedge F(T_1,C_1,S_1 \cup S_3)$ where $S_3$ is the set of literals of the remaining 
variables contained in $t_1$. It is not hard to see that the number of possible $S_3$
is at most $2^{p-p_1-p_2}$. It follows that the number of required EBRs is at most $2^p$, each of them
formed as the conjunction of the respective BR, available as one of outputs of $D$ and the set of at most $p$
literals requiring $O(p)$ gates for their computation. Thus we conclude that $O(2^pp)$ gates will be enough for computing  
of all the required EBRs of $F(T_1)$ and $F(T_2)$.
Summing up numbers of gates considered throughout the proof, we conclude
that $O(3^p)$ additional gates will be sufficient for our purpose. $\blacksquare$

{\bf Proof of Theorem \ref{dnnfcnf}.}
We order nodes of $T$ so that every child appears before
its parent. By induction on this order relation, we prove that it is possible to construct
a DNNF of size $O(3^p|T|)$ whose outputs compute all BRs of $F(T')$ for all the subtress
$T'$ of $T$. To make Lemma \ref{layerdnnf} working for 
the case where a non-leaf node has only one child, we extend $T$ so that such nodes have 
an additional child being a leaf node with the empty bag.

Let $T'$ be a subtree of $T$ consisting of a single node being a leaf.
The only BRs of such node are constant $true$ and $false$ functions. Thus
the number of BRs over all leaf nodes is $O(1)$, so regarding these nodes the inductive claim
holds. Applying Lemma \ref{layerdnnf} inductively, for each non-singleton subtree $T'$, we observe that
in order to compute basic residuals of $F(T')$ requires at most $3^p$ additional gates, so the claim stands
for each non-singleton subtree $T'$ as well and for $T$ in particular. It remains to compute $F(T)$.
Applying the generalized Shannon expansion, we observe that $F(T)$ is a disjunction of at most $2^p$ EBRs of
$F(T)$ , however the additional $O(2^p)$ gates preserve the asymptotic.  
The runtime of this construction is discussed in detail in the appendix. $\blacksquare$
\section{Discussion}
In this paper we presented a theorem that shows that a circuit of treewidth $k$ can be 
transformed into, roughly speaking, an equivalent circuit of treewidth $9k+2$ with at most $4$ times 
more gates. A consequence of this statement is that any space-efficient knowledge compilation
parameterized by the \emph{treewidth} of the input circuit can be transformed into a space
efficient knowledge compilation parameterized  by the \emph{cliquewidth} of the input circuit.
We elaborated this consequence on the example of {\sc dnnf}. 
As a result we obtained a theoretically efficient but formidably
looking space complexity of $(9^{18k}k^2n)$. Therefore, the first natural question is how likely it is that this huge 
exponent base can be reduced. 

The next question for further investigation is to check if the proposed upper bound can be applied to {\sc sdd} \cite{SDD} which is more 
practical than {\sc dnnf} in the sense that it allows a larger set of queries to be efficiently handled. To answer this question
positively, it will be sufficient to extend Theorem \ref{dnnfcnf} to the case of {\sc sdd}, the `upper' levels of the reasoning will
be applied analogously to the case of {\sc dnnf}. 

It is important to note that rankwidth is a better parameter for capturing dense graphs than cliquewidth
in the sense that rankwidth of a graph does not exceed its treewidth plus one \cite{RWDvsTWD} as well as cliquewidth \cite{CWDApprox}, 
while cliquewidth can be exponentially larger than treewidth (and hence rankwidth) \cite{CorRo}.
Also, computing of rankwidth, unlike cliquewidth, is known to be FPT \cite{RWDCompute}. Therefore, it is interesting to 
investigate the relationship between rankwidth and treewidth of a Boolean function. For this purpose rankwidth has to be extended
to directed graphs \cite{DirRWD}. It is worth saying that if the question is answered negatively, i.e. that treewidth
of a circuit can be exponentially larger than its rankwidth, it would be an interesting circuit complexity result.

Finally, recall that all the upper bounds on the {\sc dnnf} size obtained in this paper are polynomial in the \emph{size} of the circuit
that can be much larger than the number of variables. On the other hand, the upper bound on the {\sc dnnf} size 
parameterized by the treewidth of the primal graph of the given {\sc cnf} is polynomial in the number of variables \cite{DarwicheJACM}. 
Can we do the same in the circuit case? 
\bibliographystyle{plain}
\bibliography{KnowComp}
\appendix
\section{Cliquewidth vs.simplified cliquewidth}
To define cliquewidth, we introduce a graph $G=(V,E,L)$ where $L$ is a function
from $V$ to the set of natural numbers. 
To distinguish from the labeled graph in the {\sc scd}, we can call such graph
\emph{numerically labeled}.

We introduce the following operations on graphs.
\begin{itemize}
\item Let $i$ be a number that  $L(u) \neq i$ for all $u \in V(G)$
and let $v \notin V(G)$. The operation $i(v)$ adds a new vertex $v$ to the graph 
and extends $L$ so that the corresponding number of $v$ is $i$. 
\item Let $i,j$ be two numbers having non-empty preimages in $L$.
Then $\eta_{i,j}$ adds all possible edges between vertices labeled with $i$
and vertices labeled with $j$.
\item The operation $\rho_{i,j}$ changes to $j$ all vertices having label $i$.
\item Let $G_1=(V_1,E_1,L_1)$ and $G_2=(V_2,E_2,L_2)$ be two graphs with disjoint
sets of vertices. Then the result of disjoint union $G_1 \oplus G_2$ is the
graph $G=(V_1 \cup V_2, E_1 \cup E_2,L_1 \cup L_2)$.
\end{itemize}

A clique decomposition is a binary rooted tree $T$ every node of which is 
associated with a numerically labeled $G(t)$ graph and the following rules are observed.
\begin{itemize}
\item Each leaf node is associated with a single vertex graph.
\item Let $t$ be a node having the only child $t_1$. Then $G(t)$ is obtained
from $G(t_1)$ by one of the first $3$ operations in the above list.
\item If $t$ is a binary node with children $t_1$ and $t_2$ then $G(t)=G(t_1) \oplus G(t_2)$.
\end{itemize}

The width of the given clique decomposition is the smallest $k$ such that the images
of all vertices of all graphs $G(t)$ are members of $[1, \dots,k]$.
The cliquewidth of the given graph $G$ is the smallest $k$ such that there is a clique
decomposition $T$ with the root $r$ such that $G=(V(G(r)),E(G(r)))$, i.e. the function
$L(G(r))$ may be arbitrary.

For the rest of the discussion we need to choose sutiable terminology.
First, abusing the notation, we associate the decompositions with their trees,
especally as the function $G(t)$ allows to obtain the graph associated with 
a particular node. Let $G$ be a numerically labeled graph. Then $G'=Lb(G)$ 
is a labeled graph such that the elements of ${\bf S}(G')$ are sets of vertices assigned
with the same number by $L$. Let us call the number of images of the elements of the
numerically labeled graph $G$ the \emph{width} of $G$. Finally for a rooted tree $T$ we denote by
$r(T)$ the root of $T$.   

\begin{lemma} \label{widthrelation}
For any clique decomposition $T$ there is an {\sc scd} $T^s$ such that 
$Lb(G(r(T))=G(r(T^s))$ and the width of $T^s$ is at most twice larger
than the width of $T$.
\end{lemma}

{\bf Proof.}
The proof is by induction on the height of $T$. If $T$ is a leaf with the only
node associated with a graph $G$ then we create a single-node {\sc scd} 
associated with $Lb(G)$. Otherwise, assume that $r(T)$ has the only child
$r_1$ ad let $T_1$ be the subtree (and the respective clique decomposition)
rooted by $r_1$. By the induction assumption, there is $T^s_1$ satisfying
the statement of the lemma. If the operation associated with $r(T)$ is $i(v)$
then $T^s$ is a tree such that $r(T^s)$ has the only one child $r'_1$ rooting
subtree $T^s_1$ and $G(r(T^s))$ is obtained from $G(r'_1)$ by adding a new
vertex $v$. If the operation associated with $r(T)$ is $\eta_{i,j}$ then
$G(r(T^s))$ is obtained from $G(r'_1)$ by the adding new edges operation
between labels $S_i$ and $S_j$ consisting of vertices labeled by $i$ and $j$,
respectively, in $G(r_1)$. If the operation associated with $r$ is $\rho_{i,j}$
then two sitations are possible. In the first situation, graph $G(r_1)$ has 
vertices labeled by $i$ and $j$. Let $S_i$ and $S_j$ be the sets of vertices of
$G(r_1)$ labeled by $i$ and $j$, respectively. Then $G(r(T^s))$ is obtained from
$G(r'_1)$ by the union of labels $S_i$ and $S_j$. Otherwise, $T^s=T^s_1$, that is,
we even do not add a new vertex. A direct inspection shows that the lemma holds
in all the considered cases. 

Assume now that $r(T)$ is a binary node and let $r_1$ and $r_2$ be the children of
$r(T)$ and let $T_1$ and $T_2$ be the subtrees of $T$ rooted by $r_1$ and $r_2$,
respectively. By the induction assumption, there are trees (and the respective {\sc scd}s)
$T^s_1$ and $T^s_2$ satisfying the conditions of the lemma. Let 
$G'=Lb(G(r(T^s_1))) \cup Lb(G(r(T^s_1)))$ (the union operation applies to he sets of vertices,
of edges, and of labels) and let $G''=Lb(G(r))$. By definition of the involved operations,
there are ${\bf S_1} \subseteq {\bf S}(G_1)$, ${\bf S_2}\subseteq {\bf S}(G_2)$ and one-to-one
correspondence $f$ from ${\bf S_1}$ to ${\bf S_2}$ such that 
$${\bf S}(G'')=({\bf S}(G') \setminus ({\bf S_1} \cup {\bf S_2}))\cup \{S \cup f(S)|S \in {\bf S_1}\}$$

Let ${\bf S_1}=\{S_1, \dots, S_x\}$. Then $T^s$, in addition to $T^s_1$ and $T^s_2$ contains nodes
$z_0 \dots,z_x$ such that the children of $z_0$ are $r(T^s_1)$ and $r(T^s_2)$ and, for for each
$1 \leq i \leq x$, the child of $z_i$ is $z_{i-1}$. Furthermore, $G(z_0)=G'$ and for each 
$1 \leq i \leq x$, $G(z_i)$ is obtained from $G(z_{i-1})$ by the union of labels $S_i$ and $f(S_i)$.
Clearly, the widh of $G'$ is at most twice larger than the width of $G(r(T))$ and he width of the
rest of the additional nodes of $T_s$ is smaller than the treewidth of $G''$. Finally,it is not hard
to see that $G(z_x)=G''$. Thus the lemma holds for the considered case. $\blacksquare$

That the {\sc scw} of a graph $G$ is at most larger than the cliquewidth of $G$ immediately
follows from Lemma \ref{widthrelation} 
\section{Runtime for Theorem \ref{finalwidth}}
\subsection{Data structure for clique decomposition}
The above approach to define the {\sc scd} is convenient for our reasoning, however the explicit
representation $(T,{\bf G})$ is too time consuming as input for an algorithm. Instead, each node
of the tree can be associated with the respective operation with pointers to labels required to
perform the operation, thus requiring a constant memory per node of $T$. 


It is not hard to see by an inductive argumentation that any two elements in ${\bf S}$ are either disjoint or one 
is a subset of the other. The labels are naturally organized into a binary tree according to the child-parent relation
with the singleton nodes being leaves. It is thus not hard to see that the number of labels ts at most $2n-1$.

We are going to show that the number of nodesof $T$ is $O(kn)$, where $k$ i the width of $(T,{\bf G})$.
Let $S \in {\bf S}(G(t))$ for some node $t$ of $T$. Then we say that $t$ \emph{contains}
$S$.

For each binary node $t$, let us identify one of the subtrees rooted by a child of this node 
as the \emph{left} subtree and the other one as the \emph{right} subtree. Then define a DAG $D$ on the
labels of $(T,{\bf G})$ as follows. The pair $(S_1,S_2)$ is an arc of $D$ if one of
the following conditions hold.

\begin{itemize}
\item $S_1$ is contained in the node where $S_2$ is created as a result of union of
labels or adding a new vertex operation.
\item Both $S_1$ ad $S_2$ are contained in a binary node, so that $S_1$ is contained in
the left subtree, while $S_2$ is contained in the right subtree.
\end{itemize}

\begin{lemma} \label{dsize}
Labels $S_1$ and $S_2$ are contained in the same node of $t$ if and only if either
$(S_1,S_2)$ or $(S_2,S_1)$ is an arc of $D$. 
\end{lemma}

{\bf Proof.}
By induction on the height of the node of $T$.
For a leaf this is obvious. Consider a non-leaf node $t$. If this node satisfies
one of the two conditions above, we are done. Otherwise, if $t$ is a unary node
then both $S_1$ and $S_2$ are contained in the only child of $t$, so the statement
holds by the induction assumption. If none of the above happens then $t$
is a binary node and both $S_1$ and $S_2$ are contained in a node of either in the left subtree
or in the right subtree. In any case both $S_1$ and $S_2$ are contained in a node of a smaller
height and again the induction assumption applies. $\blacksquare$

By definition of graph $D$, the in-degree of each vertex is at most $k-1$.
Since there are $O(n)$ labels, it follows that the number of arcs of $D$
is $O(kn)$. It follows from Lemma \ref{dsize} that the number of pairs of
labels contained in the same node is $O(kn)$. Consequently,the number of
new adjacency nodes is $O(kn)$. Since the number of the rest ofthe nodes
is $O(n)$, we conclude that the number of nodes of $T$ is $O(kn)$.

\subsection{The procedure}
We are going to demonstrate an $O(kn)$ time procedure that constructs a mixed (having both directed and undirected
edges) graph $H^*$ whose nodes correspond to the labels and two labels will be connected by either directed child-parent arcs
(going from the child to the parent) or undirected adjacency arcs. The size of this graph (the number of vertices plus the number of arcs)
will be $O(kn)$. Also, each label will be associated with a type ({\sc and}, {\sc or}, or unary). $F^*$ can be straightforwardly obtained from $H^*$
by simply substituting labels with suitable gates and the arcs with suitable wires as specified by the description of $F^*$, implying
the $O(kn)$ construction time for $F^*$.

Recall that for algorithmic purposes the {\sc scd} is represented as a tree whose nodes are associated with operations
with pointers to the labels. In the resulting graph the vertices will be associated with labels. Each label will be supplied
with the adjacency list specifying the parent and children and label connected by the adjacency arcs (if any).
Each label and each arc are the result of some operation. Therefore, exploring the tree in a topological
order from the leaves to the root, we will be able to reconstruct $H^*$. 

We start from the empty graph. If the currently considered
operation is adding a new vertex (gate) of $F$ then 
give the corresponding singleton label the type of this gate ({\sc and}, {\sc or}, or unary). If the operation is union of two 
labels $S_1$ and $S_2$ then introduce the child-parent arcs from $S_1$ and $S_2$ to $S_1 \cup S_2$. 
Technically this means following the pointer to the label $S_1 \cup S_2$ and adding pointers to $S_1$ and $S_2$
to the adjacency list marking them as children and, similarly, adding $S_1 \cup S_2$ as the parent to the 
adjacency lists of $S_1$ and $S_2$. Also, the type of $S_1 \cup S_2$ is as the type of $S_1$ (or of $S_2$, they are the same by definition
of type respecting clique decomposition). Accordingly, the adjacency operation results in adding the adjacency arc between 
the respective labels. Notice that the binary node of the clique decomposition tree does not introduce any changes: it requires union
of two disjoint graphs but at the time of exploration of the node, the union has been already performed because all the modifications
specified above are done on the \emph{same} graph, whose nodes are the set of labels of the {\sc scd}.

It is not hard to see from the description that the above procedure takes $O(1)$ time per node of $T$.
Since the number of nodes of the tree is $O(kn)$, the desired bounds follow. 

It is not hard to observe that the graph $H$ defined is Section \ref{twidth} is isomorphic to $H^*$ except that $H^*$ assigns types to
nodes and directions to edges. We will establish the tree decomposition of $H$ as specified in the proof of lemma \ref{widthh} following
post-order exploration of $T$ (children before the parents). The elements of bags will be represented by pointers to the corresponding labels.
The only element contained in leaf node $t$ is the vertex corresponding to the singleton label of $t$. Assume that $t$ is not a leaf node.
If the operation of $t$ is new adjacency then remove $t$ and make the parent of $t$ (if any) to be the parent of the only child of $t$
Otherwise, copy to the bag of $t$ all the elements from the bags of the children of $t$. Then, if the operation of $T$ is the union of labels $S_1$
and $S_2$, then replace the vertices corresponding to $S_1$ and $S_2$ by the vertex corresponding to $S_1 \cup S_2$. 
It only remains to replace each label by the respective gates of $F^*$. This algorithm spends $O(k)$ time per node of $T$. It follows 
that the overall time is $O(k^2n)$. 

\section{Runtime calculation for Theorem \ref{dnnfcnf}}
The desired DNNF is constructed inductively from the leaves
to the root. First, BRs for the subtrees rooted by leaves are constructed.
Construction of BRs for a non-singleton subtree (having constructed BRs
for the immediate subtrees) is done in 2 stages. First, the required
residuals of the children are produced. Then the BRs of the considered subtree
are produced as disjunctions of conjunctions of residuals of the children as
specified in Lemma \ref{globaldisj}. Having constructed all the basic residuals,
the desired output $F$ is constructed as a residual $F(T, \subseteq, \subseteq)$.

The difficulty of this construction is
finding pointers to the in-neighbors of the gate currently constructed. 
If implemented straightforwardly, the whole array of the currently existing
gates may have been searched, making the construction runtime quadratic
is the size of the {\sc dnnf} being constructed. We propose a more sophisticated
procedure based on amortized analysis that makes the runtime asymptotically
the same as the size of the resulting {\sc dnnf}. The description of the procedure
provided below is divided into $3$ subsections specifying the data structures,
computation of residuals of the given subtree having computed all the basic residuals
(including also computation of the root), and computation of the basic residuals. 
The final calculation of the runtime is given in the last subsection.

\subsection{Data structures}
The circuit is maintained in the form of adjacency list. Put it differently, there are records corresponding to each gate.
These records contain the gate and the pointers to the records of the other gates who are in and out-neighbors of
the corresponding gate of the considered record. The records are not located in a homogenouos array but rather grouped
around the nodes of the tree decomposition $(T,{\bf B})$. Let us see how to do that. 

An important subset of the gates are those whose output are BRs. 
Sligtly abusing the notation, we call these gats BRs as well. 
The pointer to each BR $F(T',C',S')$ is conatined in the
record associated with the root of $T'$. At the time of construction of the circuit, it is important to very efficiently
find the record associated with each gate of the DNNF being constructed. For this purpose, each BR $F(T',C',S')$
is associated with the elements of $C'$ and $S'$. In particular, such BRs are kept in an array, let us call it $BR(T')$.
The sets of variables and clauses of the input CNF are linearly ordered. This linear order is naturally projected to the
set of clauses and variables contained in the root $r'$ of $T'$. Denote by $Cl(r')$ and $Var(r')$ the set of clauses
and variables, respectively. Then the BRs  of $F(T')$ are put in correpondence with binary vectors indexed by $Cl(r') \cup Var(r')$.
The order of the respective coordinates is exactly as the order of the corresponding elements in the above mentioned order
of variables and clauses. Let $C$ be a clause contained in $r'$ and let $x$ be a vector corresponding to $F(T',C',S')$.
Then the coordinate of $C$ is $0$ if and only if $C \notin C'$, i.e. $C$ is not removed. If $Y$ i a variable contained in $r'$
then the coordinate of $Y$ in $x$ is $1$ if $Y \in S'$ and $0$ otherwise, i.e. $\neg Y \in S'$. The vectors $x$, considered
as binary numbers, serve as array indices. This means that $BR(T')[x]$  contains the pointer to the gate whose output 
is $F(T',C',S')$. Consequently, given $x$, this pointer can be obtained in  $O(1)$. We call $x$ the \emph{characteristic vector}
of $F(T',C',S')$.

Assume that $r'$ is not the root and let $r^*$ be the parent of $r'$. Then the storage of $T'$ also maintains 
an array $RR(T')$ of pointers to the residuals $F(T',C',S')$ such that $C' \subseteq Cl(r') \cap Cl(r^*)$ and
$Var(S') \subseteq Var(r') \cap Var(r^*)$. The vectors of $RR(T')$ are enumerated analogously to $BR(T')$
but indexed by elements of $I=(Cl(r') \cap Cl(r^*)) \cup (Var(r') \cap Var(r^*))$ ordered according to the above
mentioned order of variables and clauses of $F$.

\subsection{Construction of residuals given basic residuals}
It follows from Lemma \ref{localdisj} that each residual is a disjunction of EBRs. We are going to show how to construct
the circuit computing the residuals provided the gates computing the BRs have already been constructed. 
The first step is simple. We go along the array $RR(T')$ and 
specify in the record of each corresponding gate that this gate is a disjunction. Now we are going to create
the rest of the circuit. The first step is to create a binary vector $Pattern$ indexed by $Cl(r') \cup Var(r')$ exactly in the
same order of coordinates as for $BR(T')$ the $1$ entries correspond precisely to the elements of $I$. We also need a 
binary vector $ClVar$ indexed in the ame way the element equals one if and only if the corresponding element of
$Cl(r') \cup Val(r')$ is a clause. Both of these vectors can be prepared in a time polynomial in $k$.
Since the whole time of the computation of $RR(T')$ is exponential in $k$, this runtime may be not taken into
account, so we do not elaborate on it anymore.  Next, we process each element of $BR(T')$. The processing of the given
element $BR(T')[x]$ conists of $3$ stages.
\begin{itemize}
\item {\bf Redundancy testing.} On that stage the algorithm tests if the given BR is needed at all for the forming
of the array $RR(T')$. Let $F(T',C',S')$ be the BR of $F(T')$ corresponding to $x$. Then $F(T',C',S')$ is redundant
if and only if $C' \setminus Cl(r^*) \neq \emptyset$. Such element exists if and only if there is a coordinate
$i$ such that $x[i]=Pattern[i]=0$ and $ClVar[i]=1$. Clearly,the whole testing can be done in $O(k)$ per vector $x$
just by straightforward exploration of the vectors.   
\item {\bf Construction of the corresponding EBR.}
We specify $S'' \subseteq S'$ such that $Var(S'')=Var(S') \setminus Var(r^*)$. The elements of $S''$ correspond
to those coordinates $i$ of $x$ where $Pattern[i]=ClVar[i]=0$. Create a new conjunction with inputs being precisely
elements of $S$ and $F(T',C',S')$. This operation can be performed in $O(k)$ (we may safely assume that each coordinate
is accompanied with pointers to both of the variables). The output of the obtained conjunction is the EBR
$\bigwedge S'' \wedge F(T',C',S')$. 
\item {\bf Connecting the EBR to the input of the corresponding residual.}
We create a new vector $y$ and copy there the elements of $x$ on coordinates $i$ where $Pattern[i]=1$.
Clearly, $y$ can be created in $O(k)$. Then we connect the output of the conjunction created in the previous
item to the input of the residual pointedto by $RR[T'](y)$. This can be done in $O(1)$.     
\end{itemize}    

It follows from Lemma \ref{localdisj} and by construction that each gate pointed to by an element of
$RR(T')$ is indeed a residual of $T'$ as specified. Notice also that we have not applied the reuse of
conjnctions of literals as was specified in the proof of Lemma \ref{layerdnnf}, however, it does not
increase the asymptotic space of $O(2^k)$ nor the runtime $O(2^kk)$ spent to the contruction of residuals
of $T'$. In the case of root $r$, we need to have the residual $F(T,\emptyset,\emptyset)$ whose
output is the function of the {\sc cnf} $F$. This can be done according to the same scheme.
That is, we explore the array $BR(T)$ extracting elements $F(T,\emptyset,S)$ and forming the dijunctions
of all $\bigwedge S \wedge F(T,\emptyset,S)$.

\subsection{Construction of basic residuals}
We are now going to describe the construction of the rest of the DNNF including the gates whose outputs
are BRs, their in-neighbours and the rest of incident arcs. Let $T'$ be a subtree of $T$ having only one
node, that is its root is a leaf of $T$. In order to construct $BR(T')$, we explore all the characteristic
vectors $x$ of the basic residuals of $F(T')$. For the given $x$, $BR(T')[x]$ points to the $true$ constant
if the corresponding BR is $true$ constant and to the false constant otherwise (i.e. when the corresponding
BR is a $false$ constant). In order to keep the complexity of this step within the desired boundary, it is
essential that the CNF would be represented in the form of adjacency matrix that allows $O(1)$ testing
if the given literal belongs to a particular clause. In this case, it is not hard to see that the complexity
ofthis step is $O(2^kk^2)$. 

Assume that $T'$ contains more than one node. We assume w.l.o.g. that the root $r'$ has two children
$r_1$ and $r_2$ (the reasoning for one child is a restricted version of the reasoning for the case of
two children). Again, we explore all the characteristic vectors of $r'$. Let $x$ be such vector and
let $F(T',C',S')$ be the corresponding BR. The first step is to see if there is $C \in Cl(r) \setminus (C' \cup Cl(r_1) \cup Cl(r_2))$
such that $C$ is not satisfied by any literal of $S'$. If such $C$ is found then the respective BR is unsatisfiable
and all the algorithm has to do is to record the pointer to the $false$ constant in $BR(T')[x]$.
Otherwise, $F(T',C',S')$ is represented as the disjunction of conjunctions of pairs of residuals of $T_1$ and $T_2$.
The number of such conjnctions over all the characteristic vectors is $3^k$, hence it would not be difficult to
design a procedure whose runtime is proportional to $3^k$ multiplied by a polynomial of $k$. However, we want
to get rid of the polynomial factor and hence the procedure will be more tricky to enable the amortised analysis.

A standard data structure for amortised analysis is the binary counter. Consider a binary vector of $k$ elements
and let us compute the runtime of $2^k$ consecutive increments. Although the runtime of one particular increment
can be as large as $O(k)$ due to the carry the overall runtime is $O(2^k)$, i.e. $O(1)$ per increment.
In our construction, we use a refined version of binary counter which we call \emph{selective counter}.
In this counter, there are a number of \emph{fixed} digits and the increment is performed only on the digits
that are not fixed. There are a few ways how to keep information about non-fixed digits so that the next digit
can be found in $O(1)$. For example there may be a pointer to the rightmost non-fixed digit and each non-fixed
digit can contain a pointer to the next one and the last digit also records some bit telling the algorithm about
that.  Let $k_1$ be the number of non-fixed digits. Then, it is not hard to see that $2^{k_1}$ increments can
be performed in $O(2^{k_1})$. It is important to notice that if we use decrement instead increment then we have
the same upperbound on the runtime.

Back to the DNNF construction, given $x$, we introduce two vectors $x_1$ and $x_2$. The coordinates of $x_1$
correspond to $(Cl(r') \cap Cl(r_1)) \cup (Var(r') \cap Var(r_1))$. The element of $x_1$ whose coordinate 
corresponds to a variable $v$ equals $1$ if and only if $v \in S'$. Otherwise (i.e. if $\neg v \in S'$) the element 
equals $0$. The elements corresponding to clauses can be partitioned into the following $3$ sets.
\begin{itemize}
\item Elements, whose coordinates correspond to clauses of $C'$, equal $0$. 
\item Elements, whose coordinates coorespond to clauses of $Cl(r_1) \setminus (Cl(r_2) \cup C')$, are $1$.
\item Elements, whose coordinates correspond to clauses of $(Cl(r_1) \cap Cl(r_2)) \setminus C'$, are $0$.
\end{itemize}

The structure of vector $x_2$ is symmetric with the roles of $r_2$ and $r_1$ exchanged. The only difference is
that elements of coordinates as in the last item of the above list are $1$. 

We treat vectors $x_1$ and $x_2$ as selective binary counters with $(Cl(r_1) \cap Cl(r_2)) \setminus C'$ being
coordinates of non-fixed digits, the increment operation applied to $x_1$ and the decrement operation applied to $x_2$.
Then the algorithm proceeds as follows.
\begin{itemize}
\item We set the gate $BR(T')[x]$ points to as the OR-gate.
\item We create the data structure with two items whose initial value is $(x_1,x_2)$ perceived as
      binary vectors as defined above. The only operation of this data strcture is the modification
      applying increment to the first item and decrement to the second one.
      Let $k'$ be the number of non fixed digits in the above selective vectors.
      Then this data structure can be in $2^{k'}$ possible states including $(x_1,x_2)$ and the states obtained
      by sequences of modifications. The amortised analysis argument shows that all these states can be
      explored in $O(2^{k'})$ i.e. the time proportional to the number of states.
\item For each state $(x'_1,x'_2)$ as above, create a conjunction whose inputs will be the gates
      pointed to by $RR(T_1)[x'_1]$ and $RR(T_2)[x'_2]$ and whose output is an input of the gate OR
      as in the first item of this list.       
\end{itemize} 
It follows by construction and from Lemma \ref{globaldisj} that the output of the OR gate of $BR(T')[x]$
is indeed $F(T',C',S')$. The runtime spent to construction of $BR(T')$ can be calculated as follows.
Checking whether the given BR is a $false$ constant takes a polynomial time per $x$, so the total time 
is $2^k$ multiplied by a polynomial
of $k$. The same can be said regarding creation of the data structure as in the above list. 
Exploration of the states the data structure over all the vectors $x$ takes $O(3^k)$. This follows from
Lemma \ref{globaldisj} and from the fact that by construction, the algorithm spends $O(1)$ per such state.

The description of the procedure for creation of the {\sc dnnf} is now complete. Summarising, the runtime
calculations we see that it takes $O(3^k)$ time per node of the tree decomposition.
\end{document}